# Deep preferential evolution guided by pairwise comparisons enables motion correction in photoacoustic tomography

Karteekeya Sastry[†], Yousuf Aborahama[†], Junhao Zhu, Manxiu Cui, Lihong V. Wang*

Caltech Optical Imaging Laboratory, Andrew and Peggy Cherng Department of Medical Engineering, Department of Electrical Engineering, California Institute of Technology, 1200 East California Boulevard, Pasadena, CA 91125, USA

[†] These authors contributed equally to this work: Karteekeya Sastry, Yousuf Aborahama.

*Corresponding author: Lihong V. Wang (LVW@caltech.edu)

## Abstract

Motion artifacts in three-dimensional photoacoustic tomography (PAT), caused by extended mechanical scanning of sparse arrays, degrade image quality. Since sparse sampling results in low-quality sub-reconstructions, tracking motion directly is challenging. While final image quality could in principle guide artifact correction, conventional regularizers do not accurately reflect photoacoustic image quality. Here, we introduce deep preferential evolution (DPE), a derivative-free optimization framework where a learned image comparator guides evolutionary search over reconstruction parameters. By learning relative quality differences between same-target image pairs rather than absolute quality scores, the comparator generalized more robustly from simulation to in-vivo images (after unlabeled domain calibration) than an absolute scorer. Comparator-based DPE successfully corrected synthetic motion across diverse anatomies and backgrounds, and it mitigated artifacts from natural, unconstrained motion in human palm images. These results demonstrate that pairwise learned comparison can provide a transferable optimization objective for high-dimensional image restoration when labeled experimental training data are scarce.

## Main

Medical imaging technologies fundamentally operate on a trade-off between system cost (and complexity), acquisition time, and image quality. For instance, high-resolution three-dimensional (3D) magnetic resonance imaging (MRI) requires prolonged acquisition times that increase susceptibility to motion or necessitates expensive high-field magnets[1]. Similarly, motion-free X-ray computed tomography (CT) requires costly wide-area multi-row detector arrays along with careful radiation dose management[2]. Such a trade-off also exists in photoacoustic tomography[3] (PAT), an emerging imaging technology delivering rich optical contrast at ultrasonic resolution and depth for wide-ranging preclinical[4,5] and clinical applications[6], including breast cancer detection[7,8] and brain functional imaging[9]. In principle, because optical diffusion and acoustic propagation in a typical PA system occur in under a millisecond, single-shot 3D imaging via densely packed hemispherical ultrasonic transducer arrays[10] could yield motion-free images. However, in practice, system cost and complexity force several 3D PAT systems to mechanically scan sparse transducer arrays[11,12] to ensure adequate spatial sampling[13]. This extends acquisition times from milliseconds to several seconds or minutes, making the system vulnerable to motion artifacts that compromise image quality.

Correcting these motion artifacts is challenging because motion affects the image formation process in a complex, non-linear manner. Further, sparse high-frame-rate sub-reconstructions typically lack the image quality required to track motion directly in the image domain. When the motion is periodic or repetitive (e.g., heartbeat, respiration) and the acquisition time can be extended to span multiple cycles, time-gating (sometimes with the aid of external sensors like electrocardiograms) can isolate signals based on their motion phases, thus resulting in a sharp, motion-free reconstruction[14–16]. However, for irregular motion or limited acquisition times, the quality of the final reconstructed image becomes an important cue for guiding artifact correction. Therefore, several approaches attempt to solve a non-convex joint optimization problem, aiming to recover both the initial pressure distribution and the motion trajectory by leveraging conventional regularizers to guide the optimization towards higher quality reconstructions[17–19]. Yet, since conventional regularizers, such as total variation (TV), are not tailored to PA images, they serve as poor proxies for PA image quality and therefore have limited efficacy.

To address these challenges, we introduce deep preferential evolution (DPE), a novel optimization framework that combines learned image quality evaluation with evolutionary search[20] to enable derivative-free optimization[21] of image quality over high-dimensional, coupled parameter spaces. DPE is conceptually inspired by manual parameter tuning, a routine practice in imaging where human experts iteratively vary a few (preferably weakly coupled) reconstruction parameters and perform sequential visual comparisons of reconstructed image quality to reach the optimal parameters. However, this process is subjective, non-repeatable, and limited to a few parameters. DPE automates and massively scales this process by replacing the human observer with a neural network that evaluates the quality of multiple candidate reconstructions in parallel and employs evolutionary strategies to efficiently traverse the solution space.

While surrogate-assisted evolutionary computation[22] and preference-based optimization[23] are established techniques, their application in imaging, particularly 3D PAT, is typically impeded by the scarcity of experimental training datasets. To overcome this challenge, we draw on insights in computer vision, which show that machine learning pipelines utilizing relative comparisons result in superior performance with limited training data[24,25] and better zero-shot generalization[26] over traditional absolute prediction models. Similarly, recent findings in X-ray CT[27] and digital radiography[28] have demonstrated that subjective image quality assessment by human observers is significantly more reliable and consistent when using pairwise comparisons instead of absolute scoring. Building on these findings, we considered two paradigms of DPE: image scoring and image comparison. In the image scoring paradigm, a scorer returns a continuous scalar for an input image representing its quality, akin to no-reference image quality assessment[29,30]. In contrast, in image comparison, a comparator returns a label indicating which of two candidate reconstructions, generated from the same underlying target, exhibits superior quality. We hypothesize that when trained exclusively on simulated data, absolute scorers tend to overfit to simulation-specific features, leading to severe performance degradation when applied to experimental data. In contrast, joint comparison of same-target images enables the network to evaluate relative structural differences arising from the nature and severity of reconstruction artifacts rather than rely on absolute, domain-specific features, leading to robust simulation-to-experiment transfer.

In this paper, we introduce DPE and apply it to intra-acquisition motion correction in 3D PAT. By training exclusively on simulated PA images, we demonstrate that a comparator achieves superior generalization to in-vivo data (with unlabeled domain calibration) than a scorer. We validate this comparator-driven DPE framework for motion correction over approximately 300 coupled motion parameters across diverse anatomical targets, varying background levels, and an out-of-distribution motion model. Finally, we demonstrate DPE's practical utility by successfully reducing artifacts arising from natural, unconstrained human motion in in-vivo palm images. These results demonstrate that learned comparison of same-target image pairs traverses the simulation-to-experiment gap and provides a transferable objective for a high-dimensional image optimization problem.

## Results

### Schematic of DPE-based motion correction

The framework for DPE-based motion correction is illustrated in Fig. 1. DPE automates and scales manual parameter tuning (Fig. 1a) by replacing the human observer with a neural network that evaluates image quality to guide evolutionary search algorithms[20] over high-dimensional parameter spaces (Fig. 1b). Here, we apply DPE to correct motion artifacts in 3D PAT, where systems typically mechanically scan sparse (e.g., arc-shaped) ultrasonic transducer arrays to parsimoniously achieve hemispherical detection (Fig. 1c). This scanning extends the acquisition time, making the system vulnerable to motion. At each rotation angle, a sub-reconstruction is performed, which is typically of poor quality due to the sparse sampling. When there is no motion (top row of Fig. 1d), summing these sub-reconstructions results in a clean, motion-artifact-free image due to the linearity of PA image formation. In the presence of motion (middle row of Fig. 1d), although the sub-reconstructions do not reveal the impact of motion, the final image quality is severely degraded. Moreover, tracking motion directly from these low-quality sub-reconstructions is challenging. Instead, DPE utilizes the final reconstructed image quality as a cue to estimate a corrective motion trajectory, which is used to apply image-domain corrections to the sub-reconstructions, thereby mitigating motion artifacts and recovering image quality (bottom row of Fig. 1d).

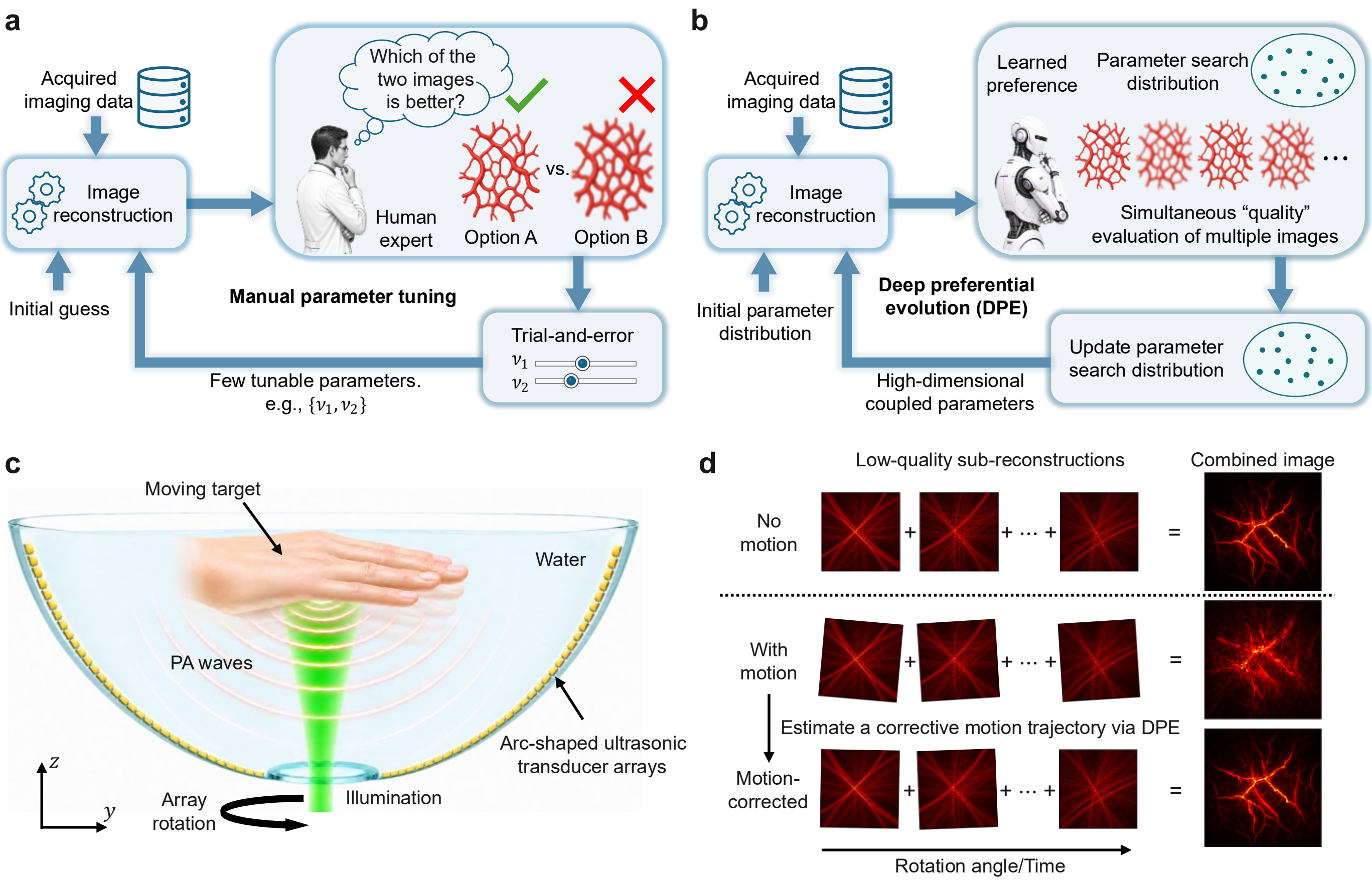

**Fig. 1| DPE framework for motion correction in 3D PAT. a,b,** Schematics of manual parameter tuning (**a**) and DPE (**b**). In manual parameter tuning, a human expert sequentially compares reconstructed images to tune a small set of parameters, starting with an initial guess, via subjective trial-and-error. DPE automates and scales this process by using a neural network to simultaneously evaluate the image quality of candidate reconstructions across a high-dimensional parameter space. Network-predicted image quality guides evolutionary search algorithms to update the parameter search distribution iteratively, starting with an initial parameter distribution, until optimal parameters are identified. **c,** Illustration of a 3D PAT system employing mechanically scanned, arc-shaped ultrasonic arrays to achieve hemispherical detection, and illumination from the bottom of the hemispherical array housing. The extended scanning makes the system vulnerable to motion. **d,** Illustration of DPE-based motion correction. Sub-reconstructions at individual rotation angles are of low quality due to the sparse transducer array. In the absence of motion (top row), summing the sub-reconstructions yields a clean, motion-artifact-free image. In the presence of motion (middle row), the final combined image quality is severely degraded. However, the low quality of the individual sub-reconstructions makes motion tracking in the image domain challenging. DPE utilizes the final image quality as a cue to estimate a corrective motion trajectory, enabling image-domain correction of the sub-reconstructions to mitigate motion artifacts and restore image quality (bottom row).

## Comparison of DPE paradigms

In PAT, where large-scale, experimental datasets are challenging to obtain, designing deep learning methodologies that train on simulations yet robustly generalize to real-world data is highly desirable. With this aim, we consider two paradigms of DPE: scoring and comparison (Fig. 2a). We hypothesize that when trained exclusively on simulated data, the comparator generalizes more robustly to out-of-distribution in-vivo data than the scorer by evaluating relative differences between images arising from the same target rather than overfitting to absolute domain-specific features. Furthermore, the comparator's task complexity is inherently lower than that of the scorer, which must learn a globally consistent quality measure across disparate targets.

To ensure a fair comparison, we trained a scorer and comparator with closely matched architectures and parameter counts (Supplementary Fig. 1) exclusively on simulated images. The training set comprised 50,000 images, generated by creating 100 synthetically motion-corrupted images for each of 500 baseline images. Each corrupted image was also assigned a ground-truth score based on the induced voxel deviation. A representative baseline simulated image, some example corrupted images with their respective voxel deviations, and the distribution of the induced root-mean-square (RMS) voxel deviation for all 100 corrupted images are displayed in Fig. 2b. Network performance was evaluated on simulated validation (40 baselines, 4,000 images) and test (8 baselines, 800 images) sets. To assess model generalization to out-of-distribution experimental data, we also had an in-vivo test set comprising 800 corrupted images, generated by synthetically inducing motion in eight in-vivo palm images of four adult human volunteers. Each of these images, acquired in 5 seconds, was visually judged to be of high quality and assumed to be motion-free. Simulated data were standardized using training statistics, whereas unlabeled domain calibration was performed for in-vivo images using statistics computed from images acquired from two independent subjects, not otherwise included in this study. The sizes of these datasets are summarized in Fig. 2c. The dataset generation and experimental data acquisition procedures are described in Methods.

We train both networks using a similar number of training examples. Their respective training pipelines are illustrated in Figs. 2d and 2e. The scorer is trained over 50 epochs, with each epoch comprising a single pass through the training set (in batches of 50 images). Random noise is added to the images before data augmentation is performed using random flips and rotations (in multiples of 90°) to double the training data size. The comparator is trained over 100 epochs, where each epoch involves a single pass over all 500 baseline vessels in the training set. Within each epoch, 12 corrupted image pairs are randomly selected for each baseline vessel, and binary comparison labels are assigned to them based on their scores. Noise is added, followed by data augmentation using identical random flips and rotations applied to both images in a pair. To encourage antisymmetry, in a second augmentation step, the order of the image pairs is swapped, and their binary labels are reversed. The scorer and comparator are trained by minimizing the mean squared error (MSE) loss and the binary cross-entropy (BCE) loss, respectively. The resulting training and validation loss curves are shown in Figs. 2f and 2g, with the model achieving the lowest validation loss selected as the final model in each case.

We evaluate the performance of the trained models on the simulated and in-vivo test sets. We plot the predicted scores against the true scores for the simulated test set in Fig. 2h. The predicted scores are in close agreement with the true scores, yielding an $R^2$ of 0.90. To assess the predicted score monotonicity (critical for optimization), we compared the predicted scores for all 4950 image pairs per baseline image and computed the comparison accuracy for each baseline by comparing the predicted comparison outputs with the true ones (top panel of Fig. 2i). In the same plot, we also obtain the comparator outputs on all (ordered) image pairs and plot the comparison accuracies. Both networks achieve high simulation accuracies of 91.3±1.9% (mean±standard deviation; sample size: 8) and 89.3±1.8%, respectively. However, on the in-vivo test set (bottom panel of Fig. 2i), the scorer's accuracy drops by 15.8 percentage points (pp) to 75.5±3.3%, whereas the comparator's accuracy only drops by 7.4 pp to 81.9±2.0%, thus demonstrating its superior generalization to out-of-distribution experimental targets (paired-sample t-test, $p = 0.004$, sample size: 8). Calculating a more conservative subject-level accuracy (by averaging the accuracies for both hands of each subject) still yielded a statistically significant difference between the two models ($p = 0.024$, sample size: 4). Moreover, the comparator outperforms the scorer on each of the 8 in-vivo examples (Supplementary Table 1). We visualized the performance differences between the two paradigms using average comparison matrices in Figs. 2j and 2k, respectively, with each panel showing the simulated (left) and in-vivo (right) test set matrices. We also quantified the near-tie comparison accuracy, swap error, and 3-cycle frequency of the scorer and comparator over the test sets (Supplementary Fig. 2). Ablations with a Siamese scalar ranker and a mismatched-target comparator (Supplementary Fig. 3) support the hypothesis that joint processing of same-target images, rather than pairwise training alone, contributes to the comparator's superior generalization.

Next, we incorporate the scorer and comparator into DPE (see Methods) and evaluate their in-vivo motion correction performance (over ~300 parameters) by correcting synthetically induced motion to the baseline images in the in-vivo test set. Representative baseline, corrupted, scorer-corrected, and comparator-corrected images are shown in Fig. 2l. While both DPE paradigms recover key features, the comparator-corrected image is sharper and has fewer artifacts than the scorer-corrected one, as confirmed by its higher structural similarity index measure (SSIM) and Pearson's correlation coefficient (PCC) relative to the baseline (Fig. 2l). We perform this on all 8 in-vivo baseline images and compare it with motion correction using a conventional regularizer (Supplementary Fig. 4). Lastly, we compute five image quality metrics (Methods)—PCC, SSIM, peak signal-to-noise ratio (PSNR), contrast-to-noise ratio (CNR), and RMS voxel deviation—for the corrupted, scorer-corrected, and comparator-corrected images for all baselines (Supplementary Table 2), and plot their average (Fig. 2m). On all metrics, the comparator-corrected images outperform the scorer-corrected ones, demonstrating that the comparator's superior generalization translates to a robust improvement in high-dimensional motion correction.

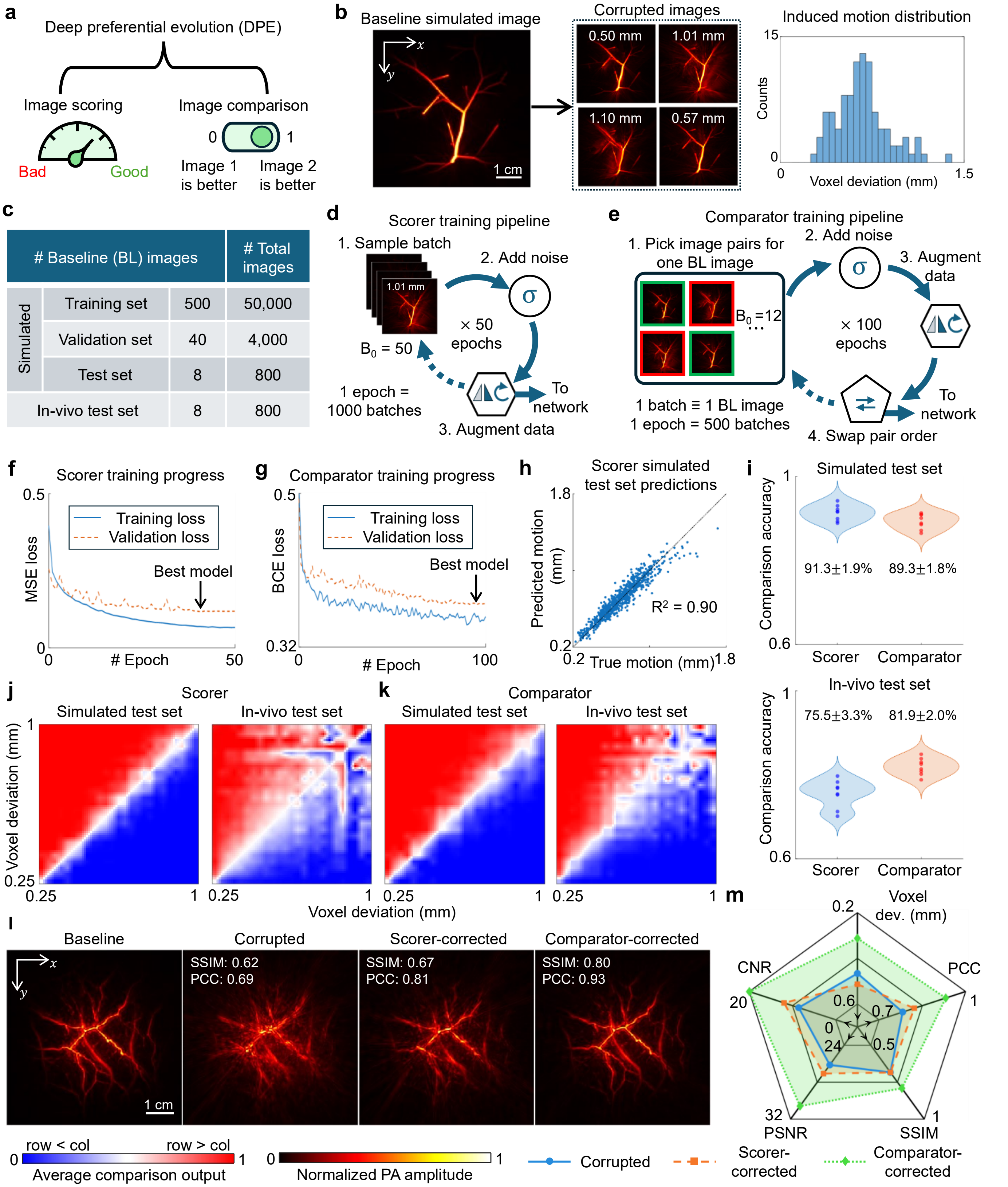

**Fig. 2 | Comparison of scoring and comparison-based paradigms of DPE. a,** Schematic of the scoring (left) and comparison (right) paradigms of DPE. The scorer maps a single image to a continuous quality metric, whereas the

comparator evaluates an image pair to identify the superior image. **b,** Synthetic training dataset generation pipeline. (Left) A representative baseline (BL) simulated vasculature image, (middle) example motion-corrupted images with their respective RMS voxel deviations, and (right) the distribution of induced RMS voxel deviations across all 100 corrupted images. **c,** Summary table of the datasets used for training and evaluating the networks. **d,e,** Schematics of the training pipelines for (**d**) the scorer and (**e**) the comparator, showing the addition of noise and training data augmentation (flips/rotations) performed for both the scorer and comparator, and the pairwise order-swapping step performed for the comparator to encourage antisymmetry. $B_0$ represents the pre-augmentation batch size for each case. The scorer and comparator have similar architectures, closely matched parameter counts, and are trained over a similar number of example images to ensure fairness in comparison. **f,g,** Training and validation loss curves for (**f**) the scorer (MSE loss) and (**g**) the comparator (BCE loss). In both cases, we select the model with the lowest validation loss. **h,** Scatter plot of the predicted versus ground-truth image quality scores on the simulated test set, demonstrating strong agreement and a high correlation ($R^2 = 0.90$). **i,** Violin plots showing the distribution of pairwise comparison accuracies across the baseline test vessels for both paradigms on the simulated test set (top, mean $\pm$ standard deviation) and the in-vivo test set (bottom), demonstrating the superior generalization of the comparator to out-of-distribution experimental data. **j,k,** Average comparison matrices for (**j**) the scorer and (**k**) the comparator on the simulated test set (left panels) and the in-vivo test set (right panels). Matrix rows (bottom-to-top) and columns (left-to-right) are ordered by increasing voxel deviation; entries are set to 0 (blue) if the row image is predicted to have lower deviation (higher quality) than the column image, and 1 (red) otherwise. **l,** In-vivo motion correction performance. From left to right: baseline palm image, corrupted image with synthetic motion, scorer-corrected image, and comparator-corrected image. All images are presented as maximum amplitude projections of their respective reconstructed 3D volumetric images along the $z$-axis. **m,** Plot of the quantitative evaluation of the corrupted, scorer-corrected, and comparator-corrected images using five averaged image quality metrics relative to the motion-free baseline for each of the 8 baseline images in the in-vivo test set: RMS voxel deviation, Pearson's correlation coefficient (PCC), structural similarity index measure (SSIM), peak signal-to-noise ratio (PSNR), and contrast-to-noise ratio (CNR).

## Effectiveness of DPE-based motion correction across anatomical targets and background levels

In the remainder of the manuscript, "DPE" will refer exclusively to the comparator-based implementation. In Fig. 3, we demonstrate that DPE is highly effective in motion correction across PA images of diverse anatomical targets—the palm, wrist/forearm, and breast—with distinct vascular structures and vessel sizes. Additionally, we show that the efficacy extends to images acquired at two different illumination wavelengths (1064 and 755 nm) that exhibit varying background levels and distinct vascular contrasts. Unlike the motion correction demonstration in Fig. 2l, where motion was introduced using the same model as the training data, here we induced smooth, temporally correlated motion to demonstrate the generalization of DPE to an out-of-distribution motion model. The baseline, synthetically motion-corrupted, and DPE-corrected images for each anatomical target and wavelength are shown in Figs. 3a–3f, respectively. Across all six cases with diverse anatomies and background levels, DPE-corrected images successfully recover the image quality, demonstrating its broad applicability despite being trained only on simulated baseline images without substantial background. To quantify this improvement, we compute the SSIM and PCC of the corrupted and corrected images relative to their baselines and present them in the top-left of the respective panels in Figs. 3a–3f. To clearly visualize the improvement, magnified views of the regions within the dashed boxes indicated in the baseline images are displayed below each image. Furthermore, we repeated the DPE-based motion correction of the corrupted images in Figs. 3a and 3b using distinct initialization seeds, demonstrating the stability of the stochastic optimization process in Supplementary Fig. 5. To validate that the optimizer corrects motion-induced artifacts rather than engaging in reward hacking by sharpening artifacts or wavelength specific features, we evaluated the cross-wavelength transferability of the estimated motion in Supplementary Fig. 6. Lastly, in Figs. 3g–3l, we plot the profiles of the images along the white dotted lines in Figs. 3a–3f, respectively.

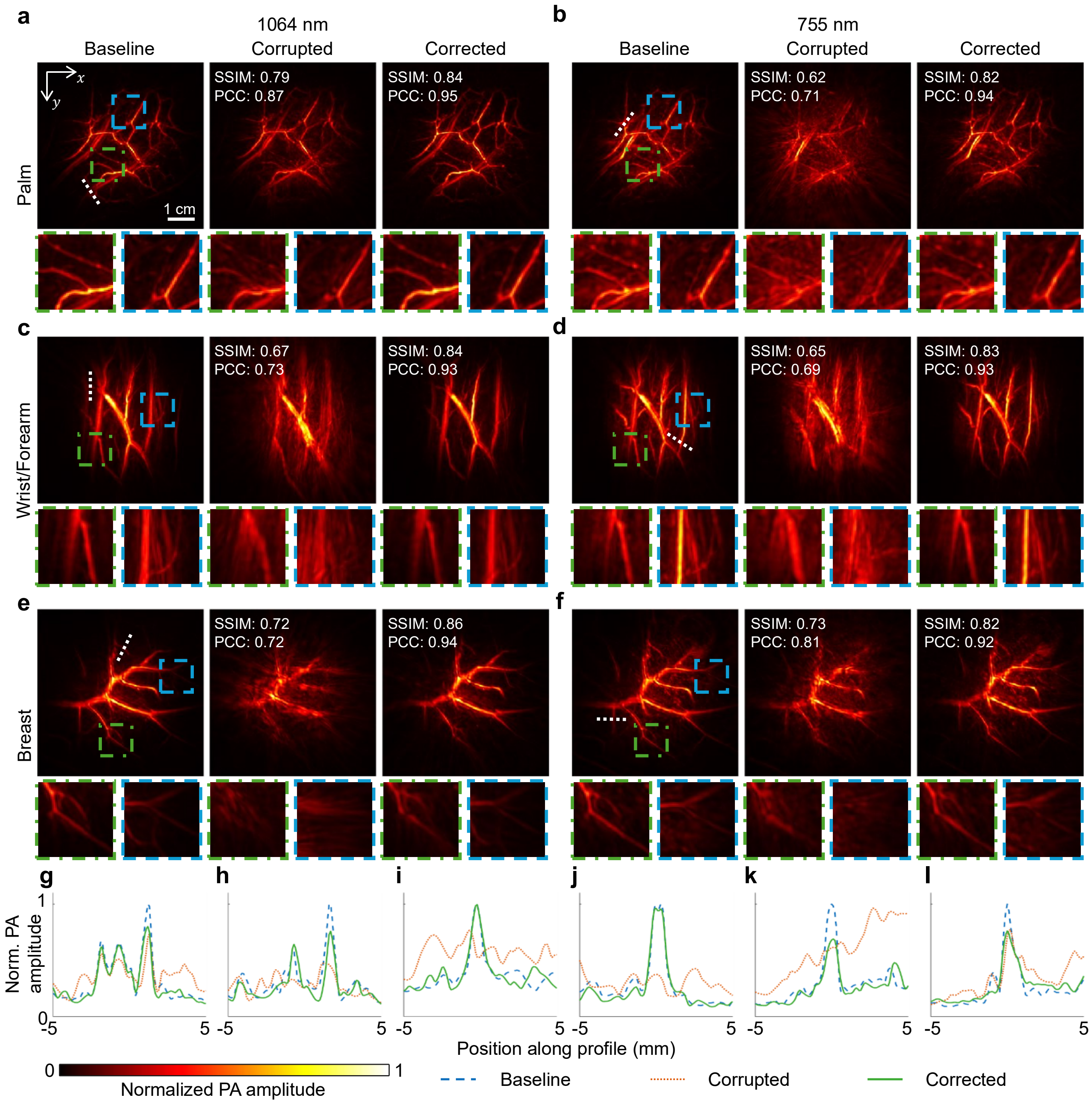

**Fig. 3 | Effectiveness of DPE-based motion correction across anatomical targets and illumination wavelengths. a–f,** Baseline (left), synthetically motion-corrupted with smooth, temporally correlated motion (center), and DPE-corrected (right) in-vivo images of the palm (**a**, **b**), wrist/forearm (**c**, **d**), and breast (**e**, **f**) acquired at illumination wavelengths of

1064 nm (**a**, **c**, **e**) and 755 nm (**b**, **d**, **f**). These results demonstrate that DPE-based motion correction performs effectively on out-of-distribution motion trajectories and across diverse anatomical targets and varying background noise levels associated with each illumination wavelength, despite the model being trained exclusively on simulated images without substantial background. To visualize the improvement clearly, magnified views of the regions indicated by the dashed green and blue boxes in the baseline panels of **a–f** are shown below each corresponding image. To quantify the improvement, the SSIM and PCC values of the corrupted and corrected images relative to the baseline are displayed in the top-left corner of each panel. All images are presented as maximum amplitude projections of their respective reconstructed 3D volumetric images along the $z$-axis. **g–l,** Profiles of the baseline, corrupted, and corrected images in **a–f**, respectively, along the dotted white lines indicated in the baseline panels.

### DPE-based motion correction of natural unconstrained human motion

While the motion correction results presented in Figs. 2 and 3 relied on synthetically induced motion, in Fig. 4, we demonstrate the performance of DPE in the motion of in-vivo images affected by natural unconstrained motion. To achieve this, we acquired palm images over extended acquisition times of up to 60 seconds while keeping the number of laser pulses (and azimuthal sampling locations) constant via downsampling (see Methods) to achieve the same signal-to-noise ratio. Figs. 4a–4d display the images of the left (Figs. 4a and 4b) and right (Figs. 4c and 4d) palms of an adult human subject, with the acquired, motion-corrupted images (pre-correction) positioned on the left side of each panel. For each palm, we present two scenarios: one with severe motion (Figs. 4a and 4c) and another with mild motion (Figs. 4b and 4d). In all four cases, the DPE-corrected images successfully recovered several fine vascular features that were blurred or obscured by motion in the raw acquisitions. To clearly visualize this improvement, we toggle between the pre- and post-correction images shown in Figs. 4a–4d and present them in Supplementary Video 1. We also visualize the progression of the optimization for Figs. 4a–4d by displaying the best-found solution at each iteration in Supplementary Video 2. Magnified views of key regions of interest, indicated by dashed squares in the pre-corrected images, are displayed below their corresponding panels, and the respective profiles along the indicated white dotted lines in Figs. 4a–4d are plotted in Figs. 4e–4h. We compare the pre- and post-correction images in Figs. 4a and 4c with their respective 5-second reference acquisitions in Supplementary Fig. 7 to validate that the recovered features are indeed true vessels. We also demonstrate the effective reduction of motion artifacts in the left and right palm images of a second human subject in Figs. 4i and 4j, respectively. We show twenty examples of pre-correction and post-correction images for the two subjects with their corresponding acquisition times and respective CNRs (see Methods) in Supplementary Fig. 8. Finally, we provide an example in Supplementary Video 3 showcasing a 5-second reference acquisition that is further enhanced by DPE-based motion correction. The consistent performance of DPE across these diverse, naturally corrupted in vivo images highlights its practical utility.

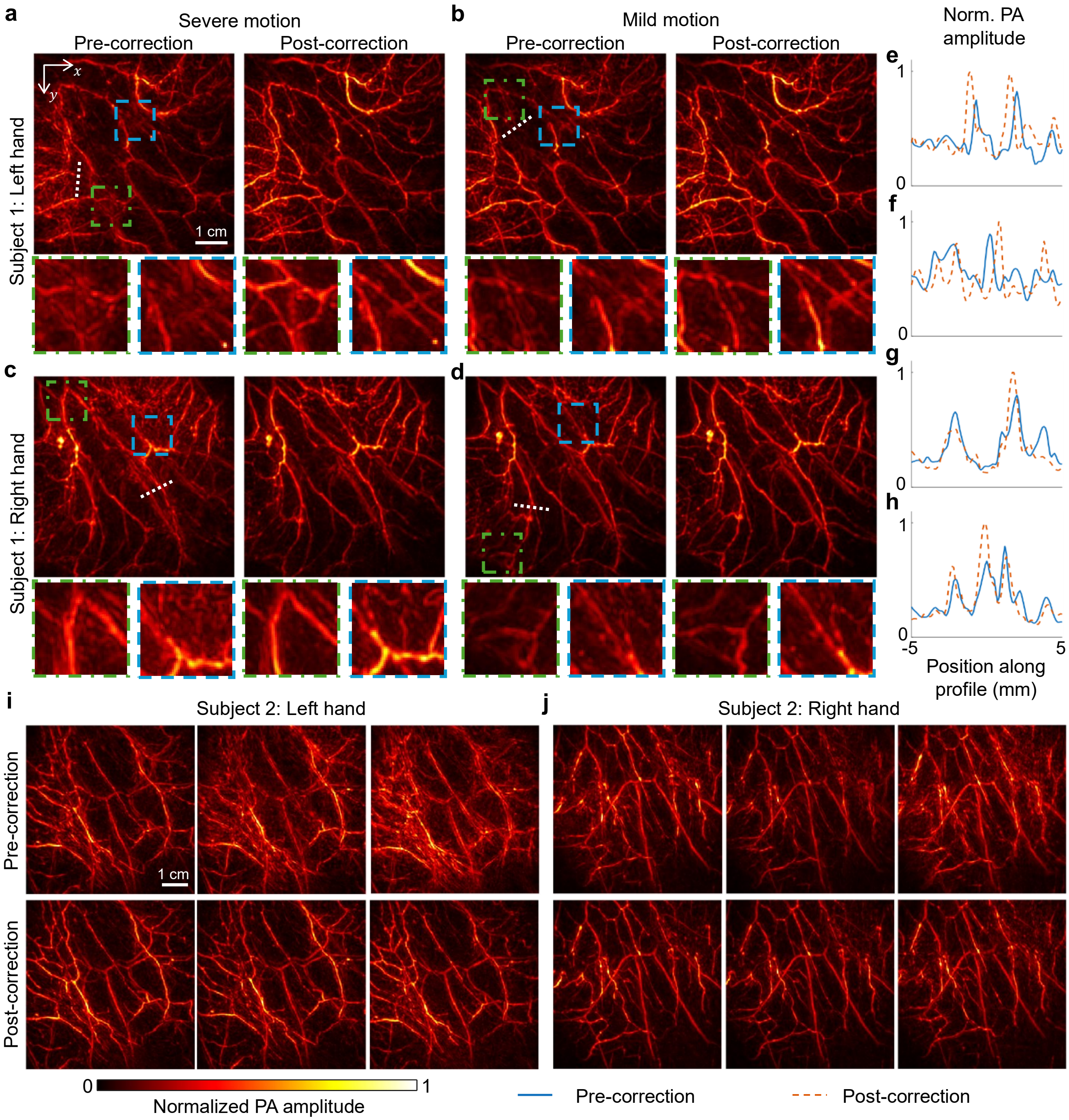

**Fig. 4 | DPE-based motion correction of in-vivo palm images with natural unconstrained motion. a–d,** Raw, motion-corrupted acquisitions (left) and corresponding DPE-corrected images (right) of the left (**a**, **b**) and right (**c**, **d**) palms of a human subject. For each palm, two scenarios are presented: severe motion (**a**, **c**) and mild motion (**b**, **d**). In all cases, DPE-based motion correction recovers key vascular features that were obscured by motion. Magnified views of the regions indicated by the dashed green and blue boxes in the baseline images of **a–d** are displayed below their respective panels. **e–h,** Line profiles of the raw and corrected images in **a–d**, respectively, taken along the dotted white lines indicated in the baseline panels. **i,j,** Raw, motion-corrupted acquisitions (top) and corresponding DPE-corrected

images (bottom) for the left (**i**) and right (**j**) palms of a second human subject. All the images in **a–d**, **i** and **j** are presented as maximum amplitude projections of their respective reconstructed 3D volumetric images along the $z$-axis.

## Discussion

Here, we introduced deep preferential evolution (DPE), a derivative-free optimization framework that couples evolutionary search with a learned image quality evaluator, and presented its application in intra-acquisition in-vivo motion correction in 3D PAT. To overcome experimental training data scarcity in PAT, we demonstrated that a comparator, which evaluates relative structural differences between images that share the same underlying target, yields robust simulation-to-experiment generalization compared to a scorer that overfits to absolute simulation-specific features. We validated that the comparator provides a domain-stable objective for intra-acquisition motion correction in 3D PAT over approximately 300 motion parameters and across varying out-of-distribution anatomical targets, background levels, and motion models. Finally, we demonstrated the practical utility of DPE by successfully mitigating artifacts in in-vivo palm images with natural unconstrained motion.

A key strength of DPE is its versatility. Beyond correcting motion artifacts across diverse anatomies and background levels, DPE could be extended to other optimization problems in PAT, where image quality is optimized by tuning underlying system or model parameters, such as aberration correction[31,32] and automated speed-of-sound tuning[33,34]. Further, the learned comparator may itself be useful beyond the trained motion correction task. In Supplementary Fig. 9, we demonstrate that the comparator trained in Fig. 2 can be applied effectively in a zero-shot manner to automated speed-of-sound tuning. Furthermore, the DPE framework may also be extended to other imaging modalities such as X-ray CT and MRI. DPE can be valuable in solving complex inverse problems where conventional physics-based approaches become computationally prohibitive or are rendered infeasible by high physical parameter uncertainty. In addition to recovering image quality, DPE relaxes the trade-off between scan time and system cost and complexity in 3D PAT. For instance, in the experimental PAT system used in this work (see Methods), quarter-arc-shaped ultrasonic transducer arrays are uniformly distributed over a hemispherical bowl that is rotated to achieve complete azimuthal sampling. Here, fewer arcs lower the overall system cost. However, fewer arcs also necessitate a larger rotation angle and, consequently, a longer acquisition time that introduces more motion artifacts. Here, DPE's motion-correction ability may enable the design of parsimonious hardware-efficient transducer arrays with minimal compromise to image quality.

It is also important to contextualize DPE against conventional deep learning approaches for medical image restoration[35], such as direct image-to-image U-Net architectures[36]. While such models can correct image artifacts in real time, they typically require extensive datasets of paired in-vivo data[37]. Moreover, direct image-domain correction carries a high risk of hallucinating seemingly realistic features[38,39]. In contrast, DPE restricts the neural network to the role of an evaluator rather than a generator. Because every candidate image is generated using the physical reconstruction model from raw measurements, the optimization is confined to a set of plausible solutions determined by the image restoration model, thus limiting the risk of artificial feature hallucination. Lastly, under severe degradation, such methods may fail to recover structures that are heavily obscured by artifacts[38].

There are a few ways to improve the current implementation of DPE. First, the execution time can be reduced. Currently, motion correction for a single in-vivo image (over 200 iterations) takes approximately 1 hour. To accelerate this, instead of relying on a global optimization method for the complete process, we can use global optimization first to get close to the optimal solution and then switch to a faster local optimization method[40,41] for the second stage. Second, as described in the Methods section, we currently assume a rigid motion model. However, for some anatomical targets, such as the human breast, incorporating a soft deformation motion model[42] may be essential for overcoming artifacts caused by physiological motion sources like heartbeat and breathing. Exploring such deformable motion models while keeping the search space dimension manageable will be an important future step. Lastly, the motion correction performance of the comparator may be improved further using optimizer-in-the-loop fine-tuning by training it on candidates that are adaptively sampled by the optimizer in addition to the randomly corrupted ones used for the initial training.

In conclusion, DPE replaces subjective manual tuning with a generalizable, simulation-trained comparator for high-dimensional coupled-parameter search. While demonstrated here for 3D PAT motion correction, DPE is broadly applicable to PAT and extendable to other imaging modalities. Ultimately, DPE may facilitate parsimonious system design by relaxing the trade-off between hardware cost, acquisition time, and image quality.

## Methods

### Experimental data acquisition

We acquired the experimental data in this work using the 3D PAT system described here[43], with a synthetic hemispherical ultrasound detection surface obtained by rotating arc-shaped arrays (Imasonic Inc.; bowl diameter: 26 cm; center frequency: 2.25 MHz; one-way 6 dB bandwidth: ~100%). We illuminated the tissue using 1064 nm light from an Nd:YAG laser (LPY7875, Litron; pulse repetition rate: 10 Hz; pulse energy: ~1.6 J; pulse width: 4–7 ns) and 755 nm light from an Alexandrite laser (Alex-Q, Beamtech Optronics Co., Ltd.; pulse repetition rate: 10 Hz; pulse energy: ~0.4 J; pulse width: 70 ns), respectively, which were expanded using an engineered diffuser (EDC-20/EDC-50, VIAVI Solutions Inc.) to be within the ANSI safety limit[44] (EDC-20, 6 cm spot size: 10.6 mJ/cm$^2$ at 1064 nm, 14.1 mJ/cm$^2$ at 755 nm; EDC-50, 10 cm spot size: 3.8 mJ/cm$^2$ at 1064 nm, 5.1 mJ/cm$^2$ at 755 nm). Unless specified otherwise, the images shown in this manuscript were acquired using 1064 nm illumination.

For the dual-wavelength demonstrations in Fig. 3, the laser pulses for the two wavelengths were interleaved to enable simultaneous acquisition of the dual-wavelength images. We used an ultrasound array along with a preamplification and data acquisition system (Photosound Inc.; maximum sampling rate: 40 MHz; dynamic range: 12 bits) to record the ultrasound waves. We streamed the recorded data to a computer using USB 3.0 and synchronized the lasers, DAQ, and scanning motor using National Instruments LabVIEW (64-bit).

**Synthetic data generation**

To generate the synthetic dataset for training the image scorer and comparator networks in Fig. 2, we started by generating 500 3D vessel images of size 256×256×128 with an isotropic voxel size of 0.25 mm using the VascuSynth package[45]. For each baseline vessel, 100 motion-corrupted images were generated assuming a rigid motion model. A synthetic hemispherical ultrasonic detection surface was defined to emulate the transducer positions of the system used for the experimental data acquisition, assuming a 5-second scan time and a 10 Hz pulse repetition rate (number of temporal frames, $N_t$ = 50). For each motion-corrupted image, a random motion trajectory describing the rigid transformation between consecutive laser pulses was generated, comprising $N_t$ 3D translation vectors $(\Delta_x, \Delta_y, \Delta_z)$ and $N_t$ 3D (Euler) rotation angles $(\alpha, \beta, \gamma)$. The translation and rotation vectors were independent and uniformly distributed within spheres of radius 0.3 mm and 0.3°, respectively.

To generate the photoacoustic signals for each of the $N_t$ laser pulses in a single acquisition, the baseline vessel volume was transformed using cubic interpolation according to the cumulative motion trajectory at each step and converted into an initial pressure map by multiplication with a static canonical 3D fluence pattern to capture the finite lateral size of the beam and exponential optical attenuation along depth. For each laser pulse, the transformed initial pressure distribution was propagated to the transducers using an analytical free-space Green's function model and a speed of sound of 1490 m/s. The propagated PA signals were convolved with a normalized first-derivative Gaussian impulse response function with a 3 MHz cutoff frequency.

The PA signals across all $N_t$ laser pulses were concatenated, and a unipolar reconstruction was obtained by applying the universal backprojection method (UBP) to the complex-valued analytical signal and taking its magnitude. The resulting motion-corrupted images had a final size of 256×256×128 with an isotropic voxel size of 0.25 mm. Lastly, a scalar motion score reflecting the extent of motion in each corrupted image was computed as the mean temporal coordinate variance across all voxels throughout the $N_t$-step trajectory. This motion score, which corresponds to the square of the RMS voxel deviation reported in Fig. 2, was found to be a good indicator of image quality, as shown in Supplementary Fig. 10. All motion-corrupted images were saved in half-precision (float16) to reduce disk storage. The same procedure was followed to generate the synthetic validation and test sets using 40 and 8 independent baseline vessels, respectively.

**Network architecture and training**

To ensure fairness in comparison, both the scorer and the comparator networks used in this work employ a similar dual-branch architecture combining spatial domain processing and frequency domain features. This dual-branch design is used because spatial convolutions excel at capturing localized structural geometry and vessel continuity, whereas Fourier domain features capture characteristic non-local phase shifts and global blur. The scorer takes as input a tensor of size (B, 1, 256, 256, 128), whereas the comparator receives a volumetric image pair of size (B, 2, 256, 256, 128), where B denotes the batch size. The main spatial branch for both networks consists of an initial 3D convolutional layer with a 3D Convolutional Block Attention Module[46] (CBAM3D; integrating channel and spatial attention), followed by five consecutive 3D residual blocks[47] with CBAM3D, and a final 3D convolutional layer, yielding a flattened spatial feature vector of size (B, 32). In parallel, the Fourier branch first applies 2 × 2 × 2 average 3D pooling to each input image to prevent aliasing and optimize GPU memory before computing the 3D Fast Fourier Transform (FFT). Then, the FFT of each image (or the concatenated FFT of an image pair) is passed through a lightweight 3D CNN comprising three convolutional layers, resulting in a flattened 16-element vector. The spatial and Fourier features are combined to give a 48-element feature vector that is finally passed through fully connected layers to produce the output score or comparison label. We use the SiLU activation function[48] in both networks and employ dropout ($p = 0.1$) to help with generalization. Ultimately, both the scorer and comparator have approximately 271,000 total parameters. Illustrations of the complete network designs are provided in Supplementary Fig. 1.

The scorer was trained for 50 epochs by minimizing the mean squared error (MSE) loss. Each epoch involved a single pass over each of the 50,000 training set images. In each batch, we randomly sampled 50 images from the training set without replacement and added zero-mean Gaussian noise to each image ( $\sigma_{\text{noise}} = 0.002 \times \max(\text{Image})$). Data augmentation was performed using random flips and rotations (in multiples of 90°), resulting in an effective training batch size of 100 images. Validation loss was computed by evaluating the network at the end of each epoch over the simulated validation set. The comparator was trained for 100 epochs by minimizing the binary cross-entropy (BCE) loss. Each epoch comprised 500 batches, where each batch corresponded to 12 corrupted image pairs from one of the 500 baseline vessels in the training set. For each pair, binary labels were assigned based on the ground-truth scores and smoothed (to 0.05 and 0.95). As with the scorer, noise was added, and data augmentation was performed using spatial flips and rotations that were applied to both images in a pair. Additionally, to encourage antisymmetry, a second augmentation step was performed in which the order of the image pairs was swapped, and the targets were inverted, yielding an effective batch size of 48 image pairs.

Validation loss was computed over deterministically selected image pairs in the validation set. All simulated images and their respective scores were standardized using statistics computed from the training dataset.

For both the scorer and the comparator, the AdamW[49] (weight decay = 0.01) optimizer was used with an initial learning rate of 0.001. A cosine annealing scheduler with a minimum learning rate of $10^{-6}$ was used to update the learning rate during training. Gradient norm clipping[50] at a maximum threshold of 1.5 was used to prevent exploding gradients during backward passes. Finally, for both networks, the model corresponding to the epoch with the lowest validation loss was selected as the final model. The scorer and comparator were trained using an NVIDIA A100 GPU (PCI-E 4.0, 40 GB HBM2 memory) on an Ubuntu 24.04.1 LTS system equipped with dual Intel Xeon 6248 R CPUs (24 cores per CPU, 3.0 GHz base, 4.0 GHz Turbo), and 12 × 128 GB DDR4-2933 ECC memory over approximately 22 hours and 51 hours, respectively.

**In-vivo image reconstruction**

For every acquired in-vivo image presented in this work, we reconstructed a complex-valued volume for each of the $N_t$ = 50 laser pulses by applying UBP to the analytical signal, yielding a total reconstruction size of 280 × 280 × 140 × 50. For the demonstration in Fig. 4, where the scanning time of the array was increased (up to 60 seconds), the measured signals were uniformly downsampled to $N_t$ = 50 azimuthal steps before reconstructing the 4D image stacks. These 4D image stacks were saved in half-precision (real and imaginary components separately) for use in the optimization procedure. For the in-vivo results in Figs. 2 and 3, the acquired/baseline images were assumed to be motion-free, and synthetic motion was added to them in the image domain to generate the motion-corrupted images. The maximum translation and rotation angle between laser pulses was set to 75% of the corresponding values used for the simulated images (0.225 mm and 0.225°, respectively), based on the empirical observation that motion beyond this threshold showed unnaturally severe motion artifacts in the in-vivo images. While in Fig. 2, the random motion was generated using the same motion model as the simulated training data, in Fig. 3, a smooth trajectory was considered. The smooth translation and rotation parameters were generated by Gaussian-filtering ($\sigma$ = 1.5) a cumulative random walk, differentiating it to generate the incremental steps, and scaling them to be within their respective maximum limits. The temporal frames of the reconstructed baseline image were transformed based on the generated motion trajectory using cubic interpolation, and the magnitude of the sum of the $N_t$ complex-valued frames was computed. The resulting image was cropped to a central size of 256 × 256 × 128 and saved along with its corresponding motion score in half-precision. All in-vivo images were standardized using statistics computed from in-vivo palm images acquired from two independent subjects, not otherwise included in this study. Nevertheless, as demonstrated in Supplementary Fig. 11, substituting this unlabeled domain calibration with self-standardization (of each volumetric image) comes at only a minor cost to overall comparison accuracy. Note that although the transducer array spatial sensitivity and a fixed illumination pattern relative to the array introduce a subtle discrepancy between applying synthetic motion to the reconstructed PA image stack (as done here) and to the underlying target (as done in simulation), we observed that this difference is negligible in this situation (and in the motion correction procedure described below).

**Deep preferential evolution (DPE) framework**

The DPE framework takes a motion-corrupted image as input and estimates a corrective motion trajectory to recover the corresponding motion-corrected image. In DPE, we formulate the motion trajectory estimation as a derivative-free optimization problem over the motion parameter search space, where we only have access to the reconstructed image corresponding to different motion parameters and the output of an image quality evaluator (via image scoring or comparison). The objective of the optimization is to find the sequence of $N_t(= 50)$ 3D translation vectors and $N_t$ rotation angles that result in the best image quality. Since the motion parameters of the first (or any single) time step only induce a global transformation for a given remaining motion trajectory, we set the parameters corresponding to the first step to 0, thus resulting in 294 parameters to be optimized.

The search space is navigated using the covariance matrix adaptation evolution strategy[51,52] (CMA-ES), an evolutionary optimization algorithm[20]. The translation and rotation variable search boundaries are fixed based on the maximum induced motion between consecutive array scanning steps, thereby implicitly enforcing a constraint on the smoothness of the trajectory. The same bounds were also applied to the natural motion demonstrations in Fig. 4, based on empirical observation of the motion severity. Optimization is performed over 200 iterations with a population size, $N_p$ = 22, starting with a mean of zero for each parameter, an identity covariance matrix, and a fixed coordinate-wise scaling that sets the initial standard deviation of each variable to one-third of its upper search bound. Active covariance matrix adaptation[53] (uses negative weights for unpromising directions), antithetic sampling[54] (reflects half of the population candidates around the mean to generate the other half), and elitism[55] (includes the best candidate solution in the next generation) are employed to accelerate convergence and retain the best-performing solution as a reference across successive generations.

In each generation in the optimization procedure, $N_p$ candidate motion trajectories are applied in reverse to generate the respective candidate volumetric images. For scorer-based DPE, a score is predicted for each candidate image and used for computing the next generation of candidate solutions. For comparator-based DPE, a tournament is conducted across all (ordered) candidate pairs in the population. The comparator outputs are recorded for every candidate pair and binarized. A fitness score is then computed for each candidate solution as the negative normalized net win-loss margin $\left(= -\frac{\text{No. of wins by a candidate - No. of losses by a candidate}}{\text{No. of comparisons involving a candidate}}\right)$ and minimized across generations. We also evaluated an alternative soft-tie handling strategy for contradictory comparator predictions in Supplementary Fig. 12 and found its performance over the in-vivo test set to be comparable with the current direct binarization strategy. The optimization was conducted using an NVIDIA A100 GPU and took 61 minutes and 66 minutes for the scorer-based and comparator-based DPE, respectively, for a single image.

**Image quality metric evaluation**

We evaluate reconstruction quality using five metrics: Pearson correlation coefficient (PCC), structural similarity index measure[56] (SSIM), peak signal-to-noise ratio[57] (PSNR), contrast-to-noise ratio (CNR), and RMS voxel deviation. PCC, SSIM, PSNR, and CNR are computed for the volumetric images after registering them to their respective baseline images. To compute PSNR and SSIM, the dynamic range was defined as the 99.9th percentile of the baseline image's voxel amplitudes to prevent outliers from skewing the metrics. The registration was performed using intensity-based automatic image registration tools in the MATLAB R2025b image processing toolbox using a rigid transformation model, with mean square error (monomodal registration) as the optimization metric and linear interpolation as the image resampling method. In Supplementary Fig. 7, the pre- and post-correction images are registered to their respective 5-second reference image. CNR is computed as $\frac{|\mu_s-\mu_{bg}|}{\sigma_{bg}}$, where $\mu_s$, $\mu_{bg}$, and $\sigma_{bg}$ denote the signal mean, background mean, and background standard deviation, respectively. To isolate the signal region, a binary mask is generated by thresholding the high-quality reference volume at 0.5 times its maximum voxel amplitude. We define the background region by dilating the binary signal mask using a 15×15×15 voxel cubic element, excluding the voxels in the signal mask, and rejecting any background voxels exceeding 0.1 times the maximum voxel amplitude of the reference. RMS voxel deviation quantifies the temporal motion trajectory error. The RMS voxel deviation of the corrupted images is computed as the square root of the average temporal variance in spatial coordinates across all voxels over the synthetically induced motion trajectory. For the motion-corrected images, we subtract the estimated motion trajectory from the induced trajectory for each voxel and compute the square root of the average temporal variance of this coordinate difference across all voxels.

**Imaging protocols**

A total of 9 adult human subjects were included in this study, with no participants excluded from the analysis. The in-vivo test set comprised 4 subjects (3 male, 1 female), and the palm and wrist dual-wavelength images belonged to subjects in this cohort. The natural motion images were acquired from 2 distinct male subjects, while unlabeled domain calibration statistics were derived from another 2 male subjects. Finally, the breast images were acquired from an independent female volunteer. All volunteers were between 18 and 65 years of age. All human imaging experiments were performed with the relevant guidelines and regulations approved by the Institutional

Review Board of the California Institute of Technology (Caltech). They were performed in a dedicated imaging room. Written informed consent was obtained from all the participants, who were made aware that their data would be deidentified in any publication, in accordance with the study protocols (IR22-1187 and IR25-0985).

**Data availability**

The acquired 4D image stacks corresponding to Figs. 4a–4d will be made publicly available upon publication. The remaining data that support the findings of this study are available within the paper and its Supplementary Information.

**Code availability**

The DPE-based motion-correction codes for the acquired images corresponding to Figs. 4a–4d, together with the trained comparator model weights, will be made publicly available upon publication. The reconstruction algorithm and other data processing methods can be found in the Methods. The reconstruction code is not publicly available because it is proprietary and is used in licensed technologies.

**Acknowledgments**

This work was supported in part by the United States National Institutes of Health (NIH) grants U01 EB029823 (BRAIN Initiative), R01 CA282505, and UG3 DA065155, and in part by grant number 2024-337784 from the Chan Zuckerberg Initiative DAF, an advised fund of the Silicon Valley Community Foundation

**Contributions**

K.S., Y.A., and L.V.W. designed the study. K.S., Y.A., and M.C. designed and developed the DPE framework. J.Z. and K.S. performed the experiments. K.S. and Y.A. analyzed the data and wrote the manuscript with input from all the authors. L.V.W. supervised the study and revised the manuscript.

**Competing interests**

L.V.W. has a financial interest in Microphotoacoustics Inc., CalPACT LLC, and Union Photoacoustic Technologies Ltd., which, however, did not support this work.

## List of supplementary materials

Supplementary Fig. 1 | Architectures of the scorer and comparator networks

Supplementary Fig. 2 | Quantitative comparison of the trained scorer and comparator

Supplementary Fig. 3 | Ablation testing of DPE paradigms for mechanistic elucidation.

Supplementary Fig. 4 | Comparison of the TV norm-corrected, scorer-corrected, and comparator-corrected in-vivo palm images

Supplementary Fig. 5 | Stability of the stochastic optimization in DPE across different initialization seeds.

Supplementary Fig. 6 | Cross-wavelength validation of DPE-based motion correction.

Supplementary Fig. 7 | Comparison of pre- and post-correction images with their corresponding 5-second reference

Supplementary Fig. 8 | Additional demonstration of DPE-based motion correction of natural, unconstrained motion

Supplementary Fig. 9 | Zero-shot application of the trained comparator to automatic speed of sound tuning

Supplementary Fig. 10 | Consistency of motion score with full-reference image quality metrics

Supplementary Fig. 11 | Comparison of unlabeled domain calibration and self-standardization on the in-vivo test set

Supplementary Fig. 12 | Comparison of direct binarization and soft-tie handling of the comparator's predictions in DPE

Supplementary Table 1 | Comparison accuracies of the scorer and comparator for each baseline vessel in the in-vivo test set.

Supplementary Table 2 | Quantitative metrics of the corrupted, scorer-corrected, and comparator-corrected images for each baseline image in the in-vivo test set.

Supplementary Video 1 | Toggling visualization of the pre- and post-correction images of Subject 1 (Figs. 4a–4d)

Supplementary Video 2 | Visualization of the progression of the optimization for the images in Figs. 4a–4d

Supplementary Video 3 | Toggling visualization of the pre- and post-correction images of an example 5-second reference image, where DPE further enhanced the image quality

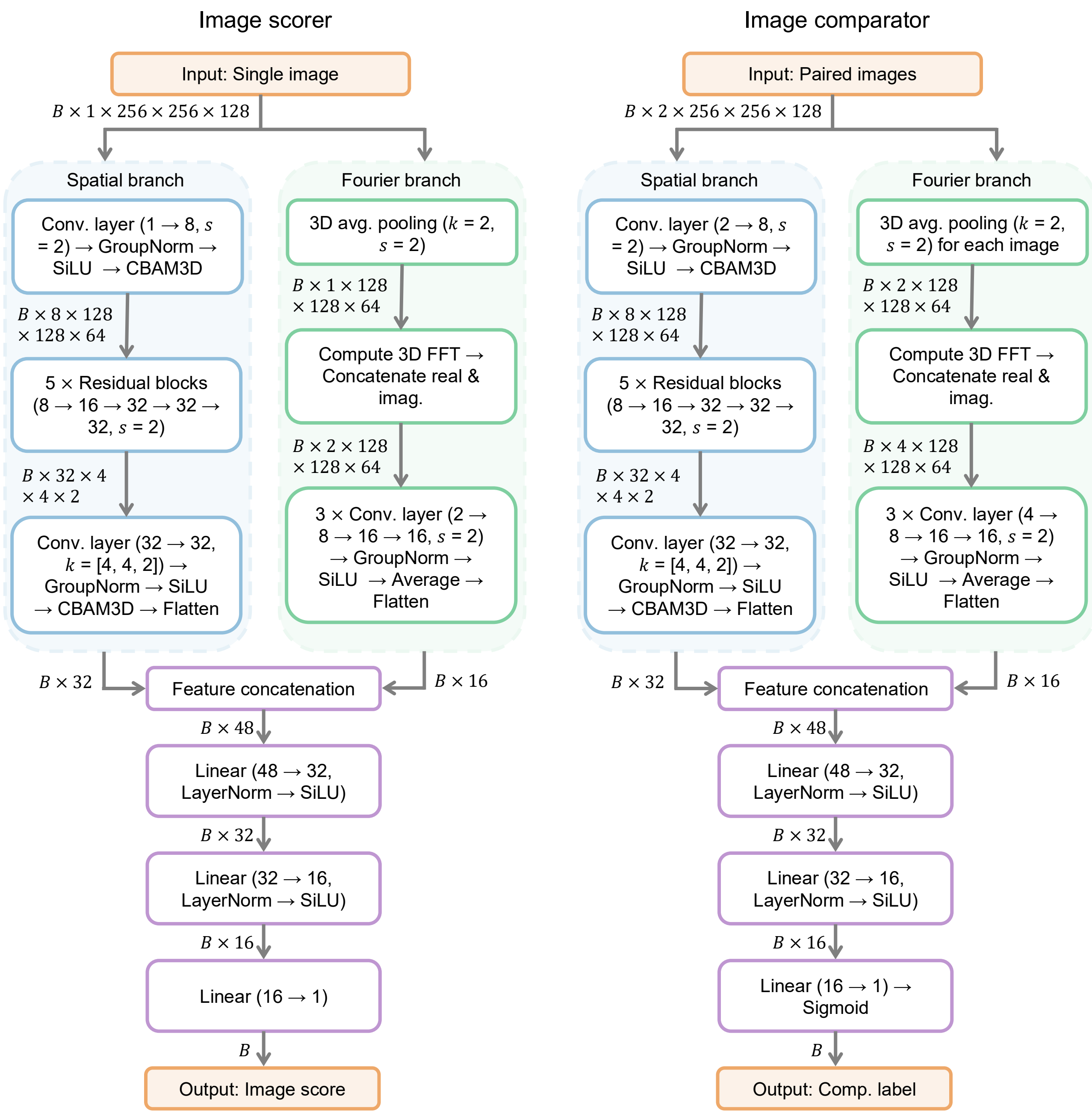

**Supplementary Fig. 1 | Architectures of the scorer and comparator networks.** Both the scorer and comparator follow a dual-branch design. The spatial branch comprises 5 residual blocks between an input and output convolutional layer. Each residual block consists of two convolutional layers, where the first convolutional layer operates with a stride ($s$) of 2, followed by a CBAM3D block that includes spatial and channel attention, a skip connection that downsamples the input by a factor of 2 before adding it to the output of the CBAM3D block, and finally a SiLU activation function. Unless specified otherwise, the convolutional layers use a kernel size ($k$) of 3×3×3. The Fourier branch first average-pools the image(s) to reduce its size, concatenates the real and imaginary parts of the FFT of the image (pair), and passes it through 3 convolutional layers. Finally, the spatial and Fourier features are flattened, concatenated, and passed through a small fully connected block that outputs the score/comparison label. B denotes the batch size.

**a**

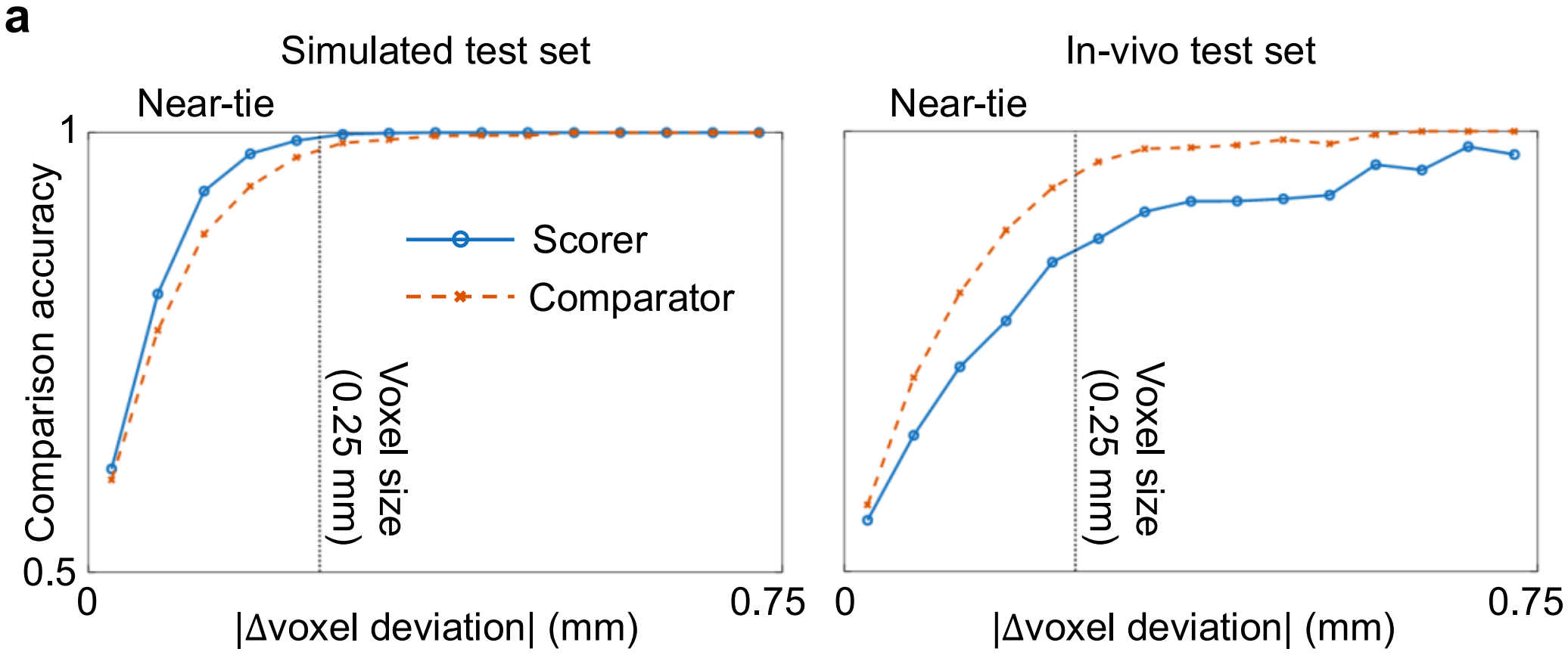

**b**

| Metric | Simulated test set | | In-vivo test set | |
|---|---|---|---|---|
| | Scorer | Comparator | Scorer | Comparator |
| Overall accuracy | 91.3% | 89.3% | 75.5% | 81.9% |
| Near-tie accuracy | 85.6% | 82.4% | 69.9% | 76.6% |
| Swap error | 0 | 0.034 | 0 | 0.106 |
| 3-cycle frequency | 0 | 0.0015 | 0 | 0.0005 |

**Supplementary Fig. 2 | Quantitative comparison of the trained scorer and comparator. a,** Plots of comparison accuracy of the scorer and comparator over the simulated (left) and in-vivo (right) test sets as a function of the absolute difference in voxel deviation (|Δvoxel deviation|). Near-tie comparisons are defined as pairs with |Δvoxel deviation| < 0.25 mm, which is the voxel size. While the scorer performs slightly better than the comparator on the simulated test set, especially in the near-tie comparisons, the comparator consistently outperforms the scorer across all |Δvoxel deviation| values on the in-vivo test set. **b,** Tabulation of the overall accuracy, near-tie accuracy, swap error, and 3-cycle frequency for the scorer and comparator over both test sets. Swap error is defined as the frequency at which the network contradicts its own predictions when the order of the input pairs is swapped. 3-cycle frequency is the proportion of circular, contradictory predictions across image triplets (e.g., the network predicts that A is better than B and B is better than C, but simultaneously also predicts that C is better than A). It is computed by exhaustively iterating over all $\binom{N}{3}$ triplets and checking whether the comparator provides circular predictions along either cyclic direction ($A \rightarrow B \rightarrow C \rightarrow A$ or $A \rightarrow C \rightarrow B \rightarrow A$). Swap error and 3-cycle frequency for the scorer are 0 by construction.

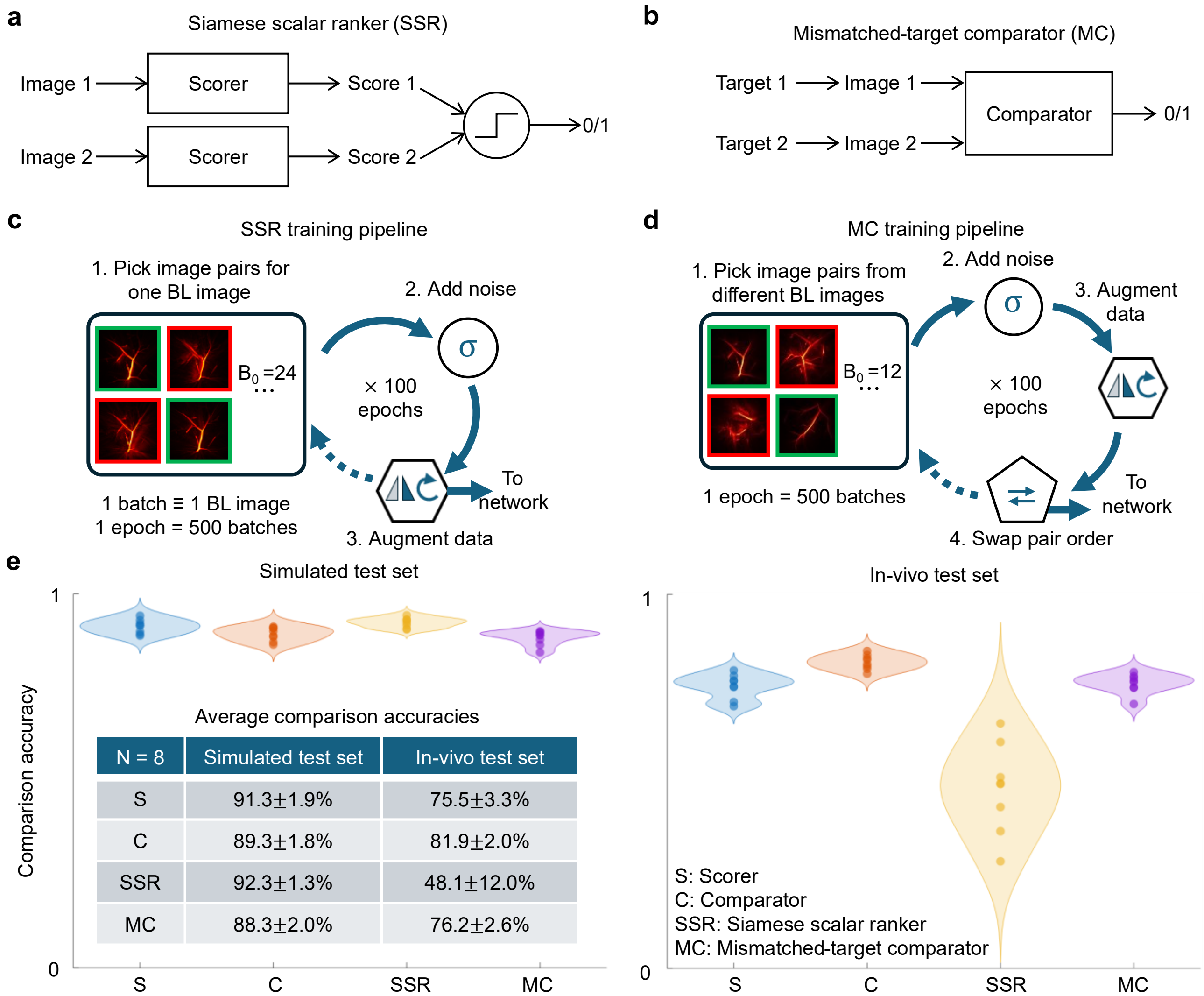

| N = 8 | Simulated test set | In-vivo test set |
|---|---|---|
| S | 91.3±1.9% | 75.5±3.3% |
| C | 89.3±1.8% | 81.9±2.0% |
| SSR | 92.3±1.3% | 48.1±12.0% |
| MC | 88.3±2.0% | 76.2±2.6% |

**f**

Simulated test set: 95% CI (percentage points) $\Delta\mu = \mu_{row} - \mu_{col}$

| | S | C | SSR | MC |
|---|---|---|---|---|
| S | | | | |
| C | [-2.8, -1.3] | | | |
| SSR | [0.3, 1.7] | [2.4, 3.8] | | |
| MC | [-4.0, -2.1] | [-2.1, 0.0] | [-5.3, -2.8] | |

**g**

Simulated test set: p-values

| | S | C | SSR | MC |
|---|---|---|---|---|
| S | | | | |
| C | < $10^{-3}$ | | | |
| SSR | 0.016 | < $10^{-3}$ | | |
| MC | < $10^{-3}$ | **0.064** | < $10^{-3}$ | |

**h**

In-vivo test set: 95% CI (percentage points) $\Delta\mu = \mu_{row} - \mu_{col}$

| | S | C | SSR | MC |
|---|---|---|---|---|
| S | | | | |
| C | [2.9, 10.0] | | | |
| SSR | [-39.1, -15.8] | [-43.4, -24.4] | | |
| MC | [-2.9, 4.4] | [-8.1, -3.3] | [17.3, 38.9] | |

**i**

In-vivo test set: p-values

| | S | C | SSR | MC |
|---|---|---|---|---|
| S | | | | |
| C | 0.004 | | | |
| SSR | < $10^{-3}$ | < $10^{-3}$ | | |
| MC | **0.655** | < $10^{-3}$ | < $10^{-3}$ | |

**Supplementary Fig. 3 | Ablation testing of DPE paradigms for mechanistic elucidation. a,b,** Illustrations of the (**a**) Siamese scalar ranker (SSR) and (**b**) mismatched-target comparator (MC). In SSR, two images are processed independently by a scorer (architecture detailed in Supplementary Fig. 1), and their output scores are compared to

generate a binary label. This framework tests whether the robust experimental generalization of the comparator is solely due to the pairwise learning structure or if it requires relative image comparison. In MC, two images, originating from different underlying targets, are processed by a comparator (architecture detailed in Supplementary Fig. 1). This tests whether the network relies on matched image pairs to compare motion artifacts, or if it is relying on absolute artifact patterns regardless of the underlying target. **c,d,** Training pipelines of SSR (**c**) and MC (**d**). Both pipelines are nearly identical to that of the comparator in Fig. 2e, except for the following modifications. Because SSR is antisymmetric by construction, we omit the additional augmentation step involving pair-order swapping. Consequently, the initial batch size, $B_0$, is doubled to 24 to maintain the same number of overall training examples. In the MC training pipeline, image pairs are generated from different underlying baseline (BL) images, training the network to evaluate images with mismatched structural anatomy. **e,** Violin plots showing the distribution of pairwise comparison accuracies for the scorer (S), comparator (C), SSR, and MC across the baseline test vessels in the simulated test set (left) and in-vivo test set (right). The average comparison accuracies (mean $\pm$ standard deviation; sample size: 8) for each case are summarized in an inset on the left. While all four models achieve high accuracy (>88%) over the simulated test set, their performance diverges significantly upon transfer to the experimental data. On the in-vivo test set, SSR's accuracy drops to a near coin-toss level of 48.1 $\pm$ 12.0%, indicating that a pairwise comparison objective alone is insufficient for robust generalization. MC, with an average accuracy of 76.2 $\pm$ 2.6%, performs on par with the scorer, but significantly underperforms relative to the comparator. This supports the hypothesis that evaluating images from the same target results in superior domain transfer due to relative feature comparison. **f–i,** Tabulation of 95% confidence intervals (CI) (**f, h**) and corresponding p-values (**g, i**) obtained from paired-sample t-tests comparing the comparison accuracies of the four paradigms on the simulated (**f, g**) and in-vivo (**h, i**) test sets. The 95% CIs represent the estimated difference in mean accuracy, defined as $\Delta\mu = \mu_{\text{row}} - \mu_{\text{col}}$ and expressed in percentage points. Statistically non-significant comparisons ($p > 0.05$) are highlighted in bold. The comparison between the scorer and comparator was our pre-planned primary hypothesis, and it is evaluated at the standard $\alpha = 0.05$ threshold. For the remaining exploratory mechanistic ablations, although the reported p-values (**g**, **i**) are unadjusted for multiple comparisons, all comparisons with $p \leq 0.05$, except the one between SSR and the scorer over the simulated test set, remain statistically significant even under a conservative Bonferroni-adjusted threshold of $\alpha = 0.0083$.

.

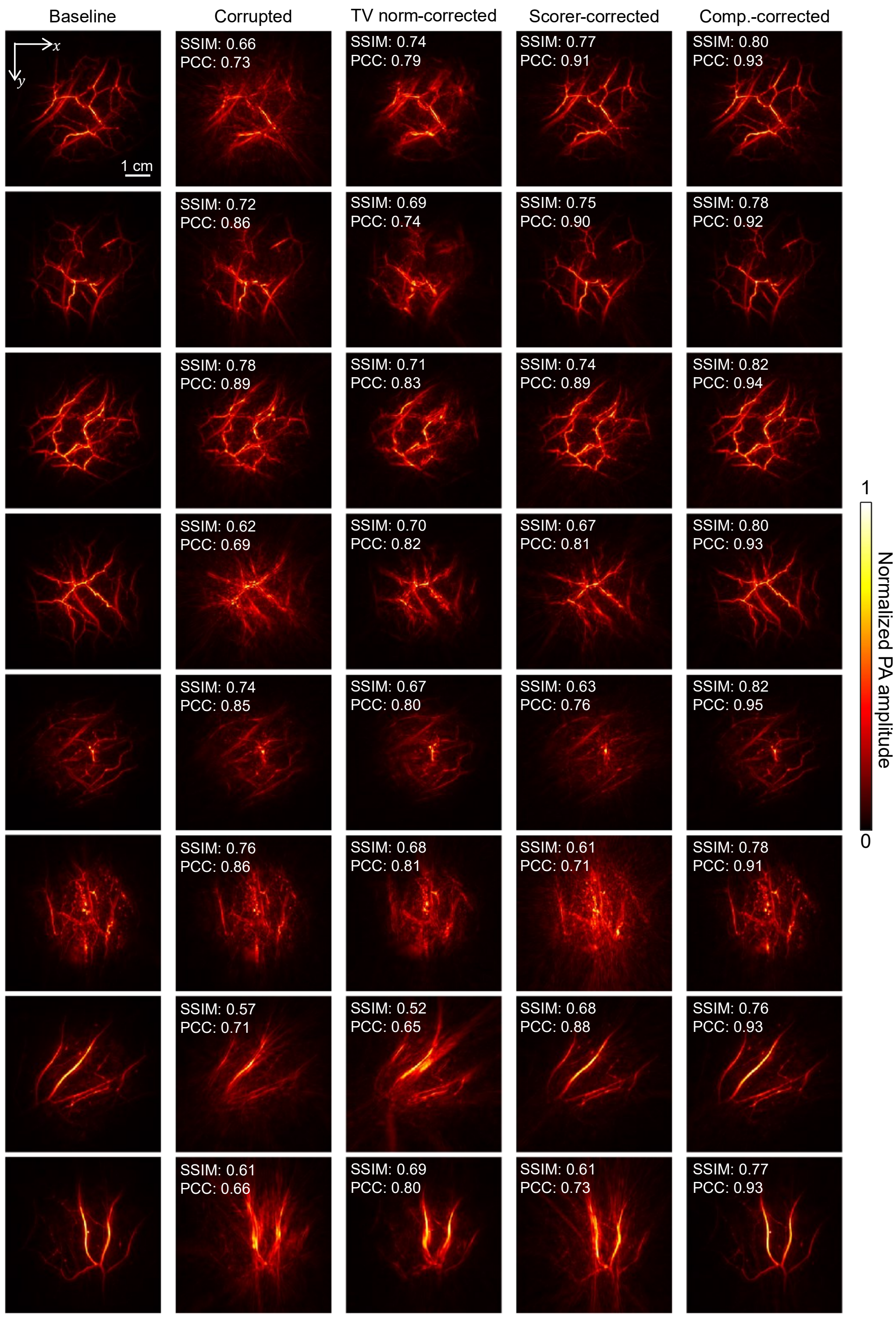

**Supplementary Fig. 4 | Comparison of the TV norm-corrected, scorer-corrected, and comparator-corrected in-vivo palm images.** Left-to-right: Baseline, synthetically motion-corrupted, TV norm-corrected, scorer-corrected, and comparator-corrected images for all 8 palm images in the in-vivo test dataset. The SSIM and PCC values of each panel with their respective baseline images are shown in the top-left of the corresponding panels. To obtain the TV-norm corrected images, we replace the scorer/comparator in DPE with TV-norm (used in prior work[18,19]) as an inverse measure of image quality that we minimize. We see that the comparator-based correction achieves superior motion artifact correction compared to the scorer-based correction as well as the TV-norm-based correction across all 8 examples.

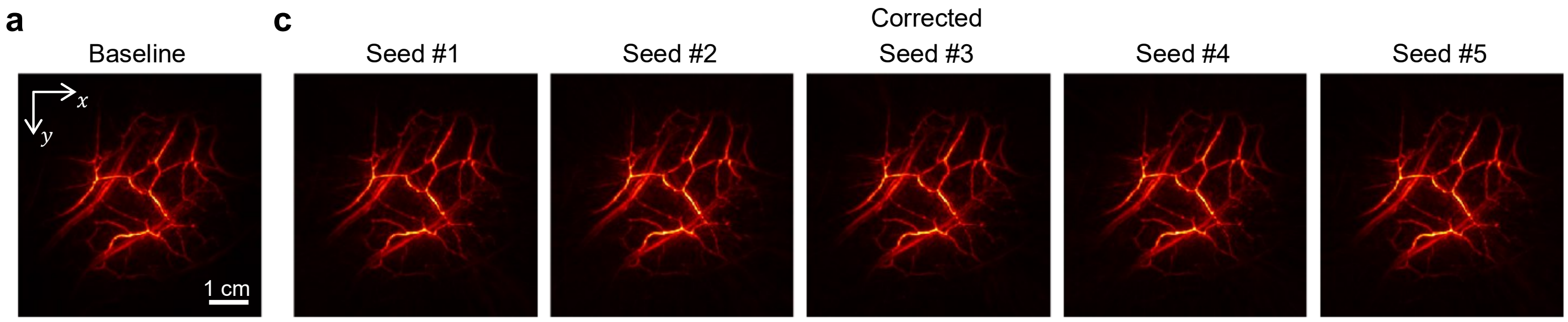

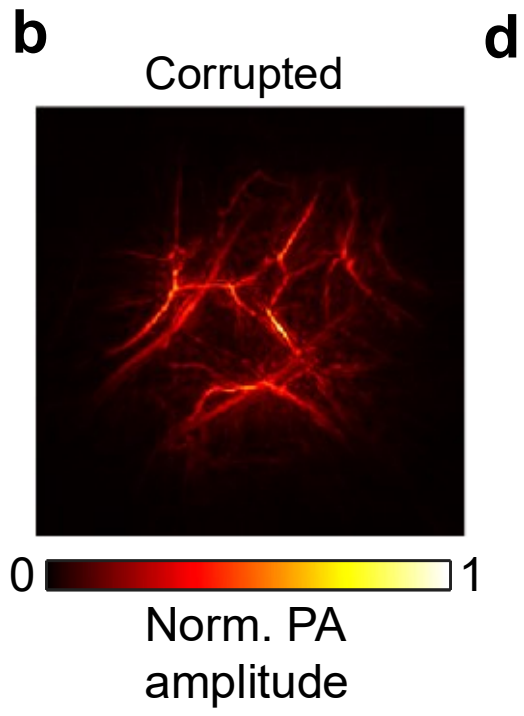

**d**

| Metric | Corrupted | Corrected | | | | |
|---|---|---|---|---|---|---|
| | | Seed #1 | Seed #2 | Seed #3 | Seed #4 | Seed #5 |
| PCC | 0.87 | 0.95 | 0.94 | 0.95 | 0.94 | 0.95 |
| SSIM | 0.79 | 0.85 | 0.82 | 0.85 | 0.83 | 0.84 |
| PSNR | 28.4 | 32.9 | 31.6 | 32.5 | 32.1 | 32.3 |
| CNR | 11.2 | 20.4 | 18.6 | 19.9 | 19.9 | 20.0 |
| Vx. dev. | 0.37 mm | 0.19 mm | 0.23 mm | 0.23 mm | 0.23 mm | 0.20 mm |

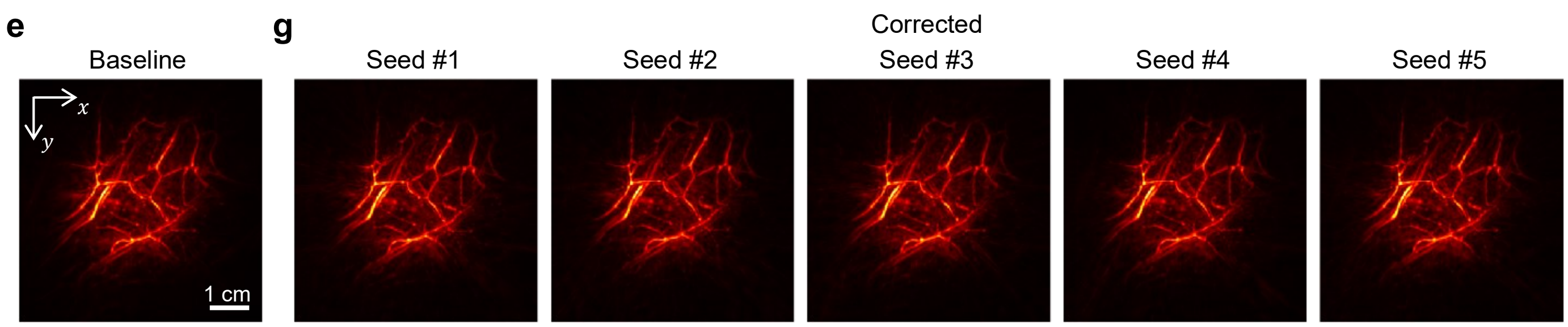

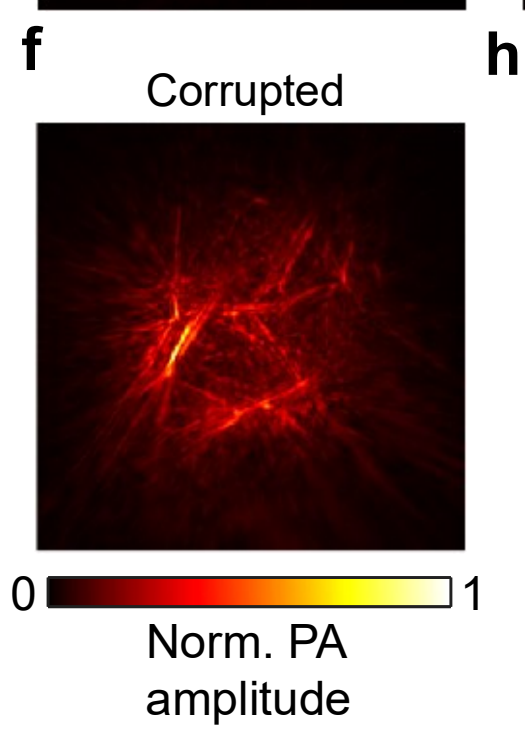

**h**

| Metric | Corrupted | Corrected | | | | |
|---|---|---|---|---|---|---|
| | | Seed #1 | Seed #2 | Seed #3 | Seed #4 | Seed #5 |
| PCC | 0.71 | 0.94 | 0.95 | 0.95 | 0.95 | 0.94 |
| SSIM | 0.62 | 0.82 | 0.84 | 0.83 | 0.83 | 0.82 |
| PSNR | 24.6 | 31.3 | 32.2 | 31.9 | 31.8 | 31.6 |
| CNR | 6.9 | 18.9 | 19.7 | 19.9 | 19.2 | 19.4 |
| Vx. dev. | 0.71 mm | 0.29 mm | 0.24 mm | 0.25 mm | 0.26 mm | 0.24 mm |

**Supplementary Fig. 5 | Stability of the stochastic optimization in DPE across different initialization seeds. a–d,** Stability analysis for the corrupted image in Fig. 3a. Baseline (**a**) and synthetically motion-corrupted (**b**) images corresponding to Fig. 3a; DPE-corrected images (**c**) for five different initialization seeds; Quantitative comparison (**d**) of the corrected images across five metrics: PCC, SSIM, PSNR, CNR, and RMS voxel deviation (mm). **e–h,** Stability analysis for the corrupted image in Fig. 3b. Baseline (**e**) and synthetically motion-corrupted (**f**) images corresponding to Fig. 3b; DPE-corrected images (**g**) for five different initialization seeds; Quantitative comparison (**h**) of the corrected images across the five metrics in **d**. For both cases, all seeds result in high-quality reconstructions that significantly reduce the artifacts in their respective corrupted images. The motion-corrected images in Figs. 3a and 3b were generated using seed #5, which was fixed prior to this analysis.

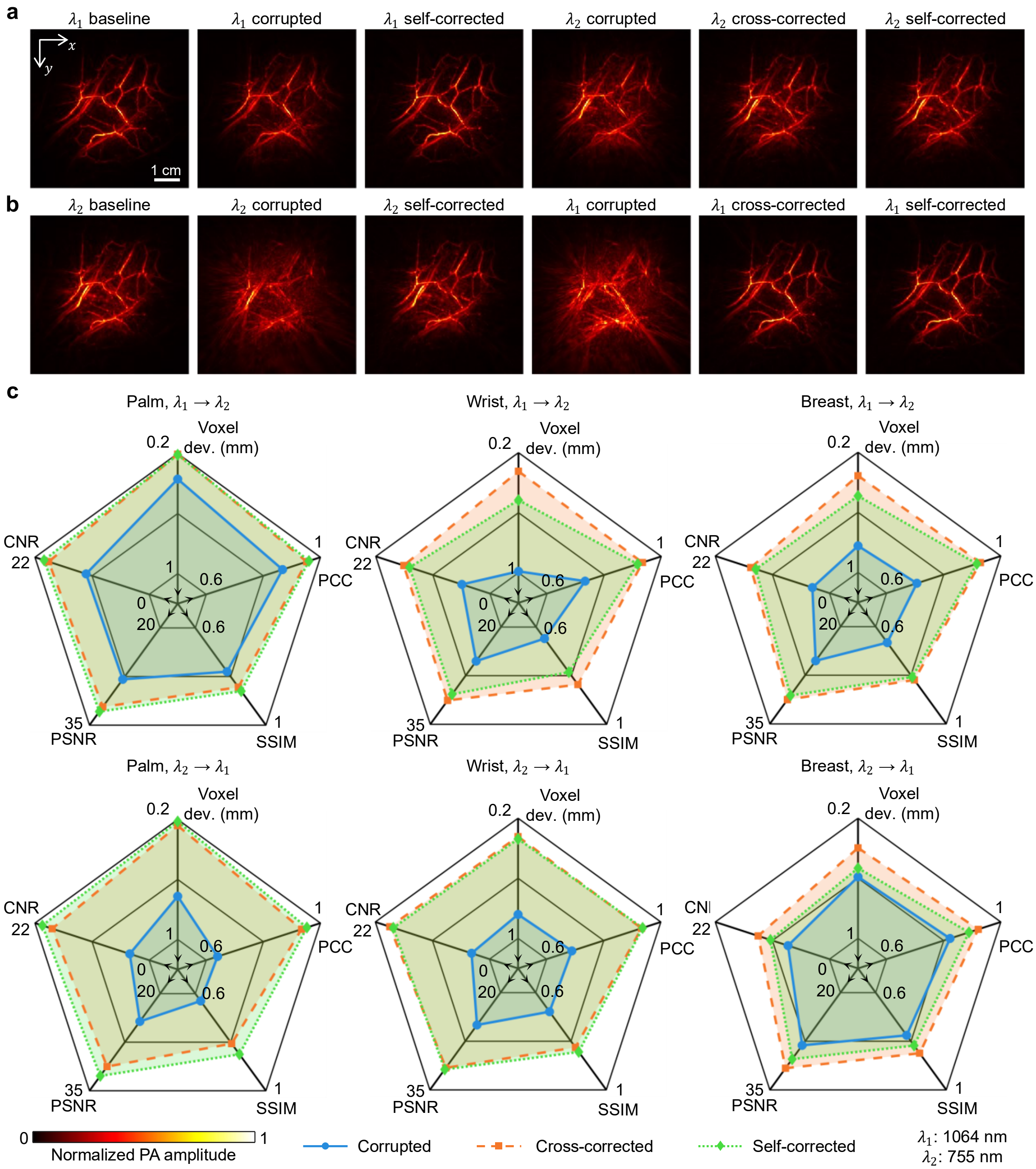

**Supplementary Fig. 6 | Cross-wavelength validation of DPE-based motion correction.** To demonstrate that DPE recovers physically meaningful motion trajectories that do not simply sharpen artifacts or wavelength-specific features, we performed cross-wavelength validation of DPE on simultaneously acquired dual-wavelength data ($\lambda_1$: 1064 nm; $\lambda_2$: 755 nm) under synthetic motion. **a,b,** In-vivo palm images under trajectory transfer from $\lambda_1 \rightarrow \lambda_2$ (**a**) and $\lambda_2 \rightarrow \lambda_1$ (**b**). The synthetic trajectory for motion corruption at the target $\lambda$ is the same as the one used for the source $\lambda$ in Fig. 3. Self-corrected denotes an image corrected using a trajectory obtained from data at the same wavelength. Cross-corrected denotes an image corrected using a trajectory obtained from data at the other wavelength. **c,** Quantitative comparison of the corrupted, cross-corrected, and self-corrected images for the different anatomical images in Fig. 3. Successful cross-wavelength transfer under matched synthetic motion indicates that the estimated corrections capture shared motion-related structure rather than only wavelength-specific image features, mitigating concerns of reward hacking.

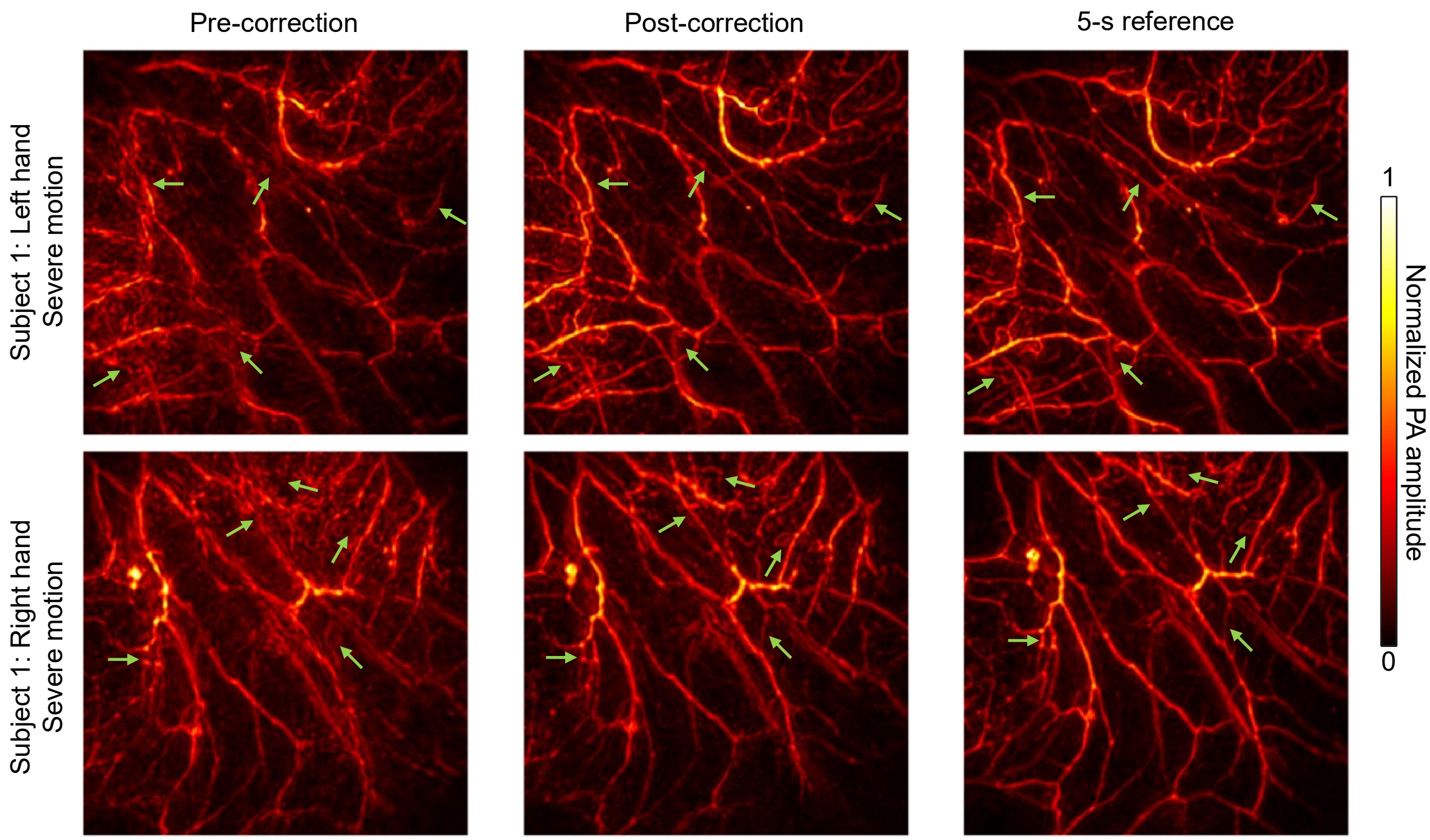

**Supplementary Fig. 7 | Comparison of pre- and post-correction images with their corresponding 5-second reference.** (Left to right) Pre-correction, post-correction, and 5-s reference images of the left and right palms of Subject 1 under severe motion (Figs. 4a and 4c). The 5-second reference image has minimal motion artifacts and serves as a validation for the vascular features recovered by DPE-based motion correction. Five representative recovered vascular features are marked by green arrows for each case.

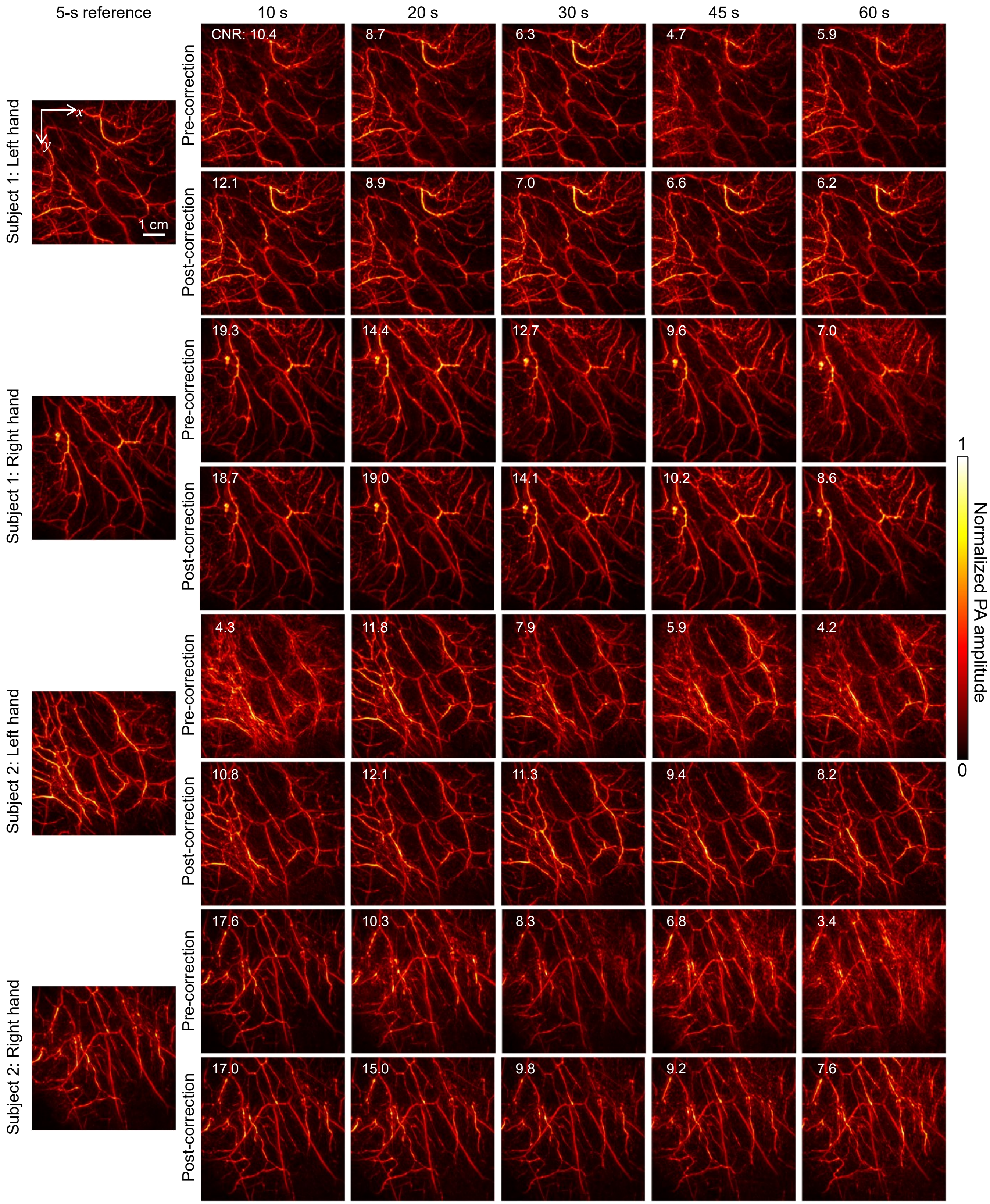

**Supplementary Fig. 8 | Additional demonstration of DPE-based motion correction of natural, unconstrained motion.** In-vivo palm images acquired over extended scanning times up to 60 seconds with natural, unconstrained

motion, along with their DPE-based motion-corrected images for the left and right hands of two adult human subjects. Reference images for each case, acquired in 5 seconds, are shown on the left. The CNR of each image is displayed in the top-left of each panel. The CNRs of almost all images increased post-correction. The only exceptions were two 10-second acquisitions, where motion was already low. This minor degradation in CNR can be partly attributed to inaccuracies in the comparator when evaluating nearly identical images, as well as uncompensated non-rigid deformations between the reference 5-second acquisition (used for the CNR computation) and the pre-corrected images (see Methods).

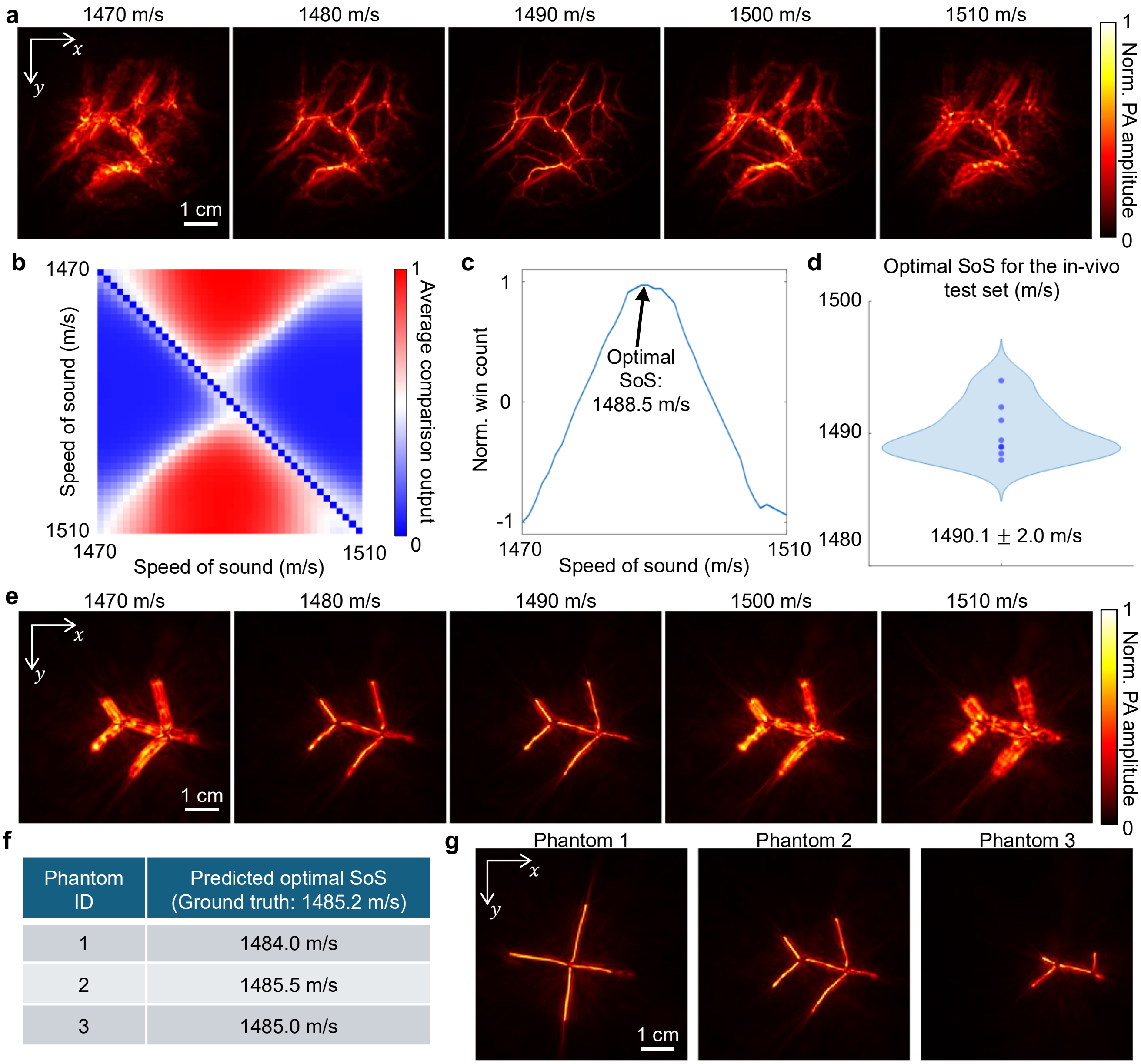

| Phantom ID | Predicted optimal SoS (Ground truth: 1485.2 m/s) |
|---|---|
| 1 | 1484.0 m/s |
| 2 | 1485.5 m/s |
| 3 | 1485.0 m/s |

**Supplementary Fig. 9 | Zero-shot application of the trained comparator to automatic speed of sound (SoS) tuning. a–d,** In-vivo SoS tuning. **a,** Reconstructed images of a representative image from the in-vivo test set for different SoS values. **b,** Matrix obtained by performing pairwise comparisons using the trained comparator from Fig. 2 on in-vivo palm images corresponding to the dataset shown in **a**, reconstructed for different SoS values between 1470 and 1510 m/s. A comparison label of 0 is given if the comparator predicts that the image corresponding to the SoS value of the row is better than that corresponding to the column, and a label of 1 is given otherwise. **c,** Plot of the normalized win count for each SoS value. The SoS that wins the most comparisons (1488.5 m/s here due to a tie) is selected as the optimal SoS value. **d,** Plot of the predicted optimal SoS values for each of the 8 images in the in-vivo test set. The comparator-predicted optimal SoS for the entire test set was 1490.1 ± 2.0 m/s (mean ± standard deviation). In contrast, a manually tuned SoS of 1490 m/s was used to reconstruct the entire dataset. **e–g,** Phantom SoS tuning demonstration with ground truth validation. **e,** Reconstructed images of a representative phantom for different SoS values. **f,** Tabulated comparison of the comparator-predicted optimal SoS against the ground-truth (1485.20 ± 0.15 m/s derived from a water temperature measurement[58] of 20.90 ± 0.05 °C) for three independent phantoms. **g,** Reconstructed images of the three phantoms using the true SoS.

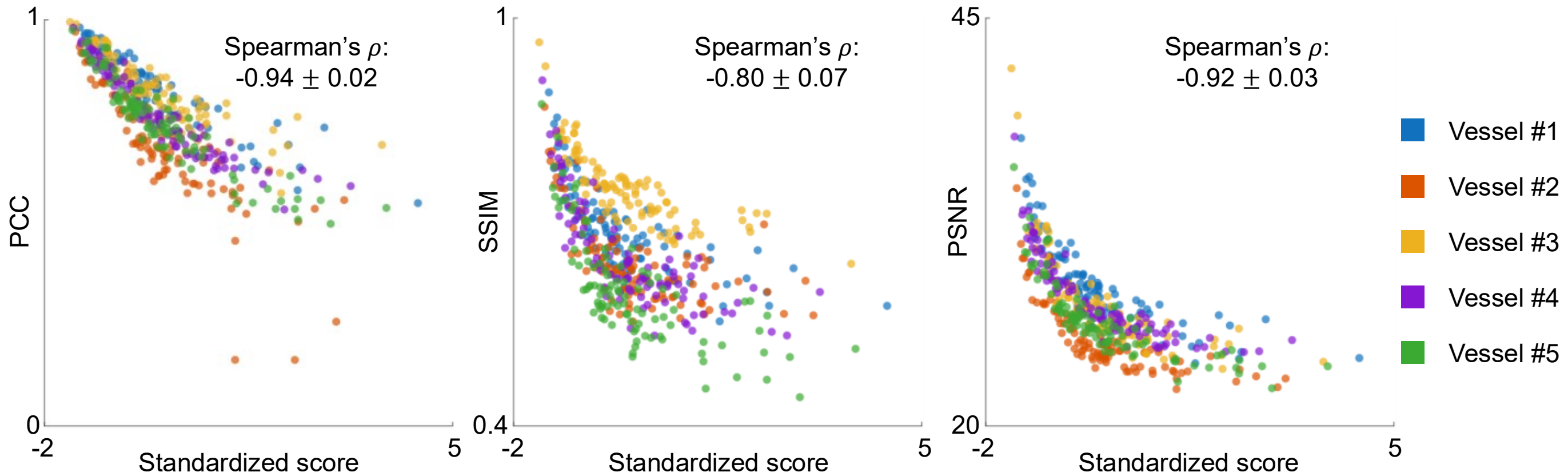

**Supplementary Fig. 10 | Consistency of motion score with full-reference image quality metrics.** Scatter plots of the PCC (left), SSIM (center), and PSNR (right) of motion-corrupted images with respect to their baseline image. Before computing the full-reference image quality metrics, image registration is performed with a motion-free reconstruction of the baseline vessel (see Methods) to correct any positional shifts in the corrupted images due to motion. Data is shown for five candidate baseline vessels in the simulated training set and is plotted against their standardized motion scores. The motion score exhibits strong negative correlations with all three full-reference image quality metrics, as confirmed by their respective Spearman's rank correlation coefficients (Spearman's $\rho$).

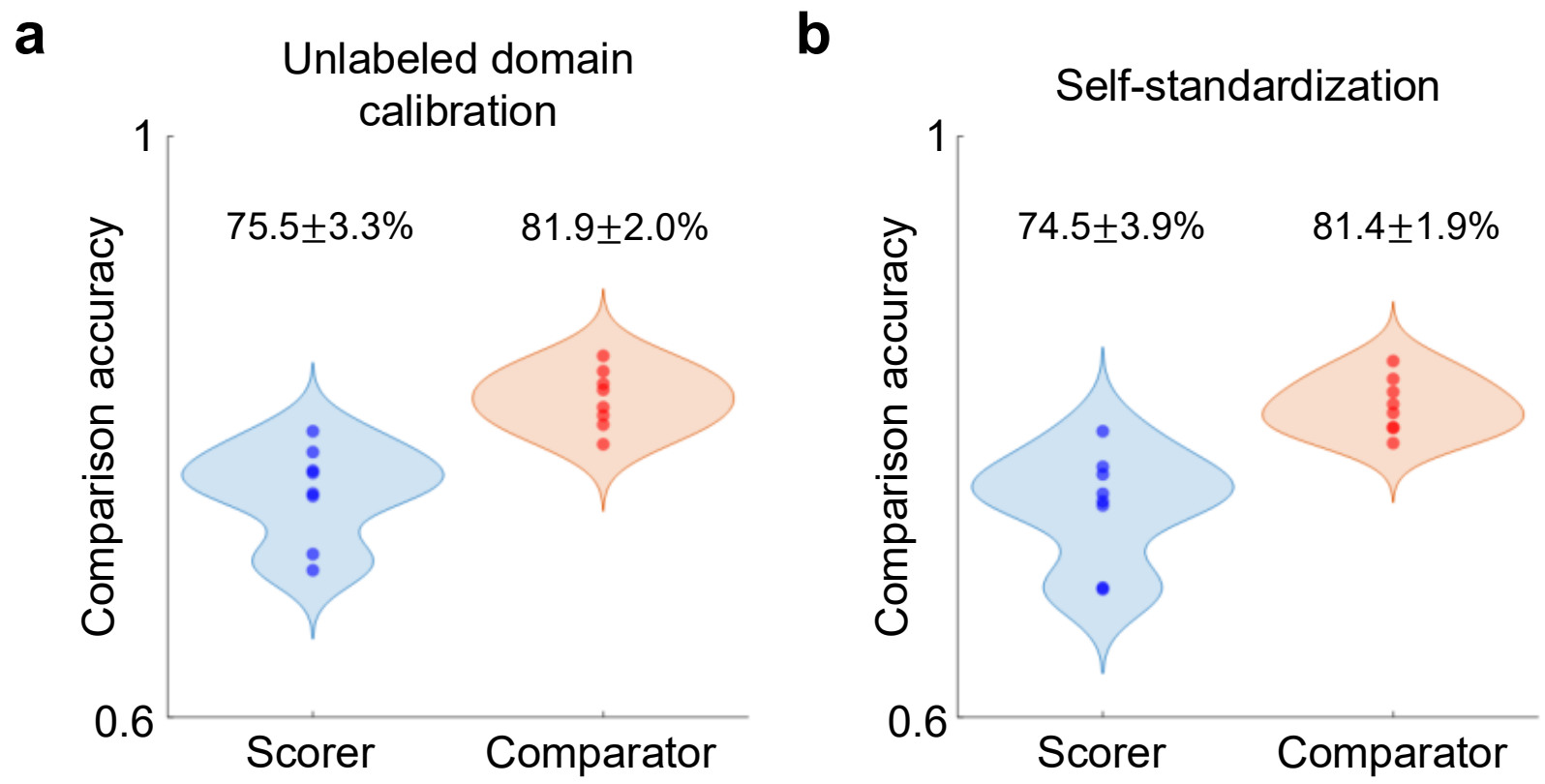

**Supplementary Fig. 11 | Comparison of unlabeled domain calibration and self-standardization on the in-vivo test set. a,b,** Violin plots showing the distribution of comparison accuracies for the scorer and the comparator across the 8 baseline vessels in the in-vivo test set using unlabeled domain calibration (**a**) and self-standardization (**b**). Unlabeled domain calibration is performed by standardizing the in-vivo images using statistics computed from in-vivo palm images acquired from two independent subjects not included in this study. Self-standardization results in only a minor drop in the average comparison accuracies (∼1 percentage point) for both the scorer and the comparator.

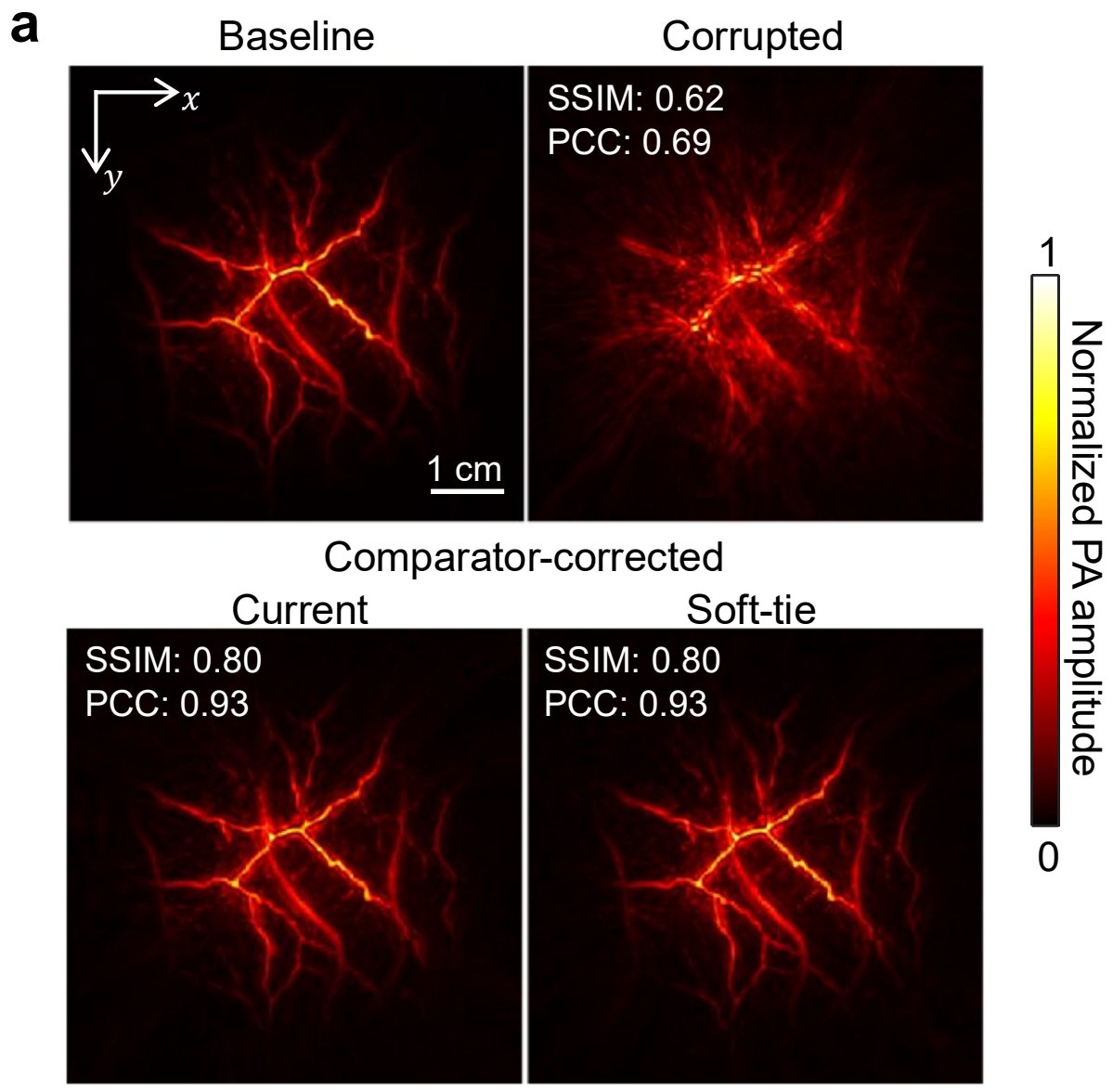

**b**

| | Corrupted | Comparator-corrected | |
|---|---|---|---|
| | | Current | Soft-tie |
| PCC | 0.78 | 0.93 | 0.93 |
| SSIM | 0.68 | 0.79 | 0.80 |
| PSNR | 26.1 | 30.6 | 30.6 |
| CNR | 8.6 | 20.0 | 19.8 |
| Voxel dev. (mm) | 0.47 | 0.31 | 0.33 |

**Supplementary Fig. 12 | Comparison of direct binarization and soft-tie handling of the comparator's predictions in DPE.** a, Representative baseline image from the in-vivo test set (top left), synthetically motion-corrupted image (top right), comparator-corrected image using the current direct binarization of the comparator's predictions (bottom left), and comparator-corrected image using a soft-tie handling strategy (bottom right). Soft-ties refer to situations where the comparator provides contradictory predictions for an input pair of images depending on their order. In these situations, we compare the comparator's predictions before binarization to break the tie. The PCC and SSIM of the corrupted and corrected images relative to the baseline are shown in the top-left of the respective panels. Visual inspection and quantitative metrics indicate that the corrected images obtained using the current direct binarization strategy and the soft-tie handling strategy are comparable. **b,** Average quantitative metrics of the corrupted images and the corrected images using the direct binarization and soft-tie handling strategies across the 8 baseline vessels in the in-vivo test set.

**Supplementary Table 1 | Comparison accuracies of the scorer and comparator for each baseline vessel in the in-vivo test set.**

| Subject # | Hand | Scorer | Comparator |
|---|---|---|---|
| 1 | Right | 71.2% | 83.8% |
| | Left | 78.3% | 83.0% |
| 2 | Right | 79.7% | 80.2% |
| | Left | 75.4% | 78.8% |
| 3 | Right | 70.1% | 81.4% |
| | Left | 77.0% | 80.8% |
| 4 | Right | 76.8% | 82.6% |
| | Left | 75.2% | 84.9% |

**Supplementary Table 2 | Quantitative metrics of the corrupted, scorer-corrected, and comparator-corrected images for each baseline image in the in-vivo test set.**

| Subject # | Hand | Corrupted | | | | |
|---|---|---|---|---|---|---|
| | | PCC | SSIM | PSNR | CNR | Vx. dev. |
| 1 | Right | 0.73 | 0.66 | 25.5 | 7.6 | 0.59 |
| | Left | 0.86 | 0.72 | 28.0 | 11.8 | 0.36 |
| 2 | Right | 0.89 | 0.78 | 29.6 | 12.6 | 0.32 |
| | Left | 0.69 | 0.62 | 24.9 | 3.6 | 0.52 |
| 3 | Right | 0.85 | 0.74 | 27.0 | 11.6 | 0.30 |
| | Left | 0.86 | 0.76 | 27.7 | 10.3 | 0.28 |
| 4 | Right | 0.71 | 0.57 | 23.7 | 7.5 | 0.61 |
| | Left | 0.66 | 0.61 | 22.7 | 3.6 | 0.74 |

| Subject # | Hand | Scorer-corrected | | | | |
|---|---|---|---|---|---|---|
| | | PCC | SSIM | PSNR | CNR | Vx. dev. |
| 1 | Right | 0.91 | 0.77 | 29.9 | 16.5 | 0.43 |
| | Left | 0.90 | 0.75 | 29.4 | 16.3 | 0.37 |
| 2 | Right | 0.89 | 0.74 | 29.4 | 14.4 | 0.44 |
| | Left | 0.81 | 0.67 | 26.8 | 9.6 | 0.46 |
| 3 | Right | 0.76 | 0.63 | 25.2 | 9.6 | 0.62 |
| | Left | 0.71 | 0.61 | 24.8 | 6.0 | 0.57 |
| 4 | Right | 0.88 | 0.68 | 27.3 | 17.3 | 0.51 |
| | Left | 0.73 | 0.61 | 23.8 | 6.0 | 0.70 |

| Subject # | Hand | Comparator-corrected | | | | |
|---|---|---|---|---|---|---|
| | | PCC | SSIM | PSNR | CNR | Vx. dev. |
| 1 | Right | 0.93 | 0.80 | 31.0 | 19.2 | 0.30 |
| | Left | 0.92 | 0.78 | 30.4 | 19.0 | 0.30 |
| 2 | Right | 0.94 | 0.82 | 32.1 | 19.2 | 0.24 |
| | Left | 0.93 | 0.80 | 31.1 | 17.7 | 0.31 |
| 3 | Right | 0.95 | 0.82 | 31.6 | 22.5 | 0.28 |
| | Left | 0.91 | 0.78 | 29.6 | 16.4 | 0.32 |
| 4 | Right | 0.93 | 0.76 | 29.3 | 22.9 | 0.37 |
| | Left | 0.93 | 0.77 | 29.5 | 22.7 | 0.39 |